\documentclass[12pt,a4,english]{article}
\usepackage[T1]{fontenc}
\usepackage[latin9]{inputenc}
\usepackage{geometry}
\usepackage{verbatim}
\usepackage{amsmath}
\usepackage{graphicx}
\usepackage{longtable}
\usepackage{setspace}
\usepackage[round,sort,authoryear]{natbib}
\usepackage{amsfonts}
\usepackage{graphicx}
\usepackage{babel}
\usepackage{amsthm}
\usepackage[toc,page]{appendix}
\usepackage{booktabs,caption}
\usepackage{enumitem}
\usepackage{amsfonts}
\usepackage{accents}
\usepackage{natbib}
\usepackage{bibentry}
\usepackage{amsmath}
\usepackage{mathtools}
\usepackage[colorlinks=true,linkcolor=blue, citecolor=blue]{hyperref}
\usepackage{pdflscape}
\usepackage{multirow}


\makeatletter
\newcommand{\nextverbatimspread}[1]{  \def\verbatim@font{    \linespread{#1}\normalfont\ttfamily    \gdef\verbatim@font{\normalfont\ttfamily}}}
\makeatother
\nobibliography*

\theoremstyle{definition}

\newtheorem{example}{Example}

\newtheorem{proposition}{Proposition}
\newtheorem{remark}{Remark}

\DeclarePairedDelimiter{\floor}{\lfloor}{\rfloor}

\numberwithin{equation}{section}

\begin{document}

\title{ {\Large Testing for Structural Change under Nonstationarity}\thanks{This work has been developed during my time as a PhD student at the Department of Economics, University of Southampton - Highfield Campus, SO17 1BJ, UK. I am grateful to my advisors Jean-Yves Pitarakis and Jose Olmo for their guidance and continuous support. I also thank Tassos Magalinos for his invaluable guidance and helpful discussions.} 
}

\author{
\\
\small{ Christis Katsouris  (PhD Candidate in Economics) }
\\
\small{Department of Economics, University of Southampton}
\\
\large{ Appendix:  Proofs of Main Results }
}
\date{\today}

\maketitle

\section{Wald OLS test for mildly integrated regressors}

We consider separately the limiting distribution of the sup Wald-OLS statistic when the regressor is assumed to be generated via a mildly integrated process. The econometric intuition in this case is that since the regressor is mildly integrated then it is expected to behave asymptotically similar to the IVX instrument. Therefore, we replace $z_t$ with $x_t$ into the corresponding sample moments and obtain the corresponding limiting terms. Moreover, intuitively in the case we have a mildly integrated regressor and since the degree of persistence is controlled by the exponent rate $\gamma \in (0,1)$, then we expect that the limiting distribution of the sup Wald-OLS statistic, when testing for an unknown break-point $\pi \in \Pi$, to weakly converge to the standard NBB. 

Consider the univariate predictive regression with multiple regressors 
\begin{align}
y_{t+1} = \left( \alpha_1 + \beta_1 x_{t} \right) I_{1t} + \left( \alpha_2 + \beta_2 x_{t} \right) I_{2t} + u_{t+1}
\end{align}

where $I_{1t} :=  \mathbf{1} \{ t \leq k  \}$ and $I_{2t} :=  \mathbf{1} \{ t > k  \}$ with $k = \floor{ T\pi}$. The set of regressors  $x_t$ is generated via the following process
\begin{align}
x_t = \left(I_p - \frac{C}{ T^{\gamma} } \right) x_{t-1} + v_{t}, \ \ \ \ \text{with} \ \ x_0 = 0. 
\end{align}
where $\gamma \in (0,1)$ is the exponent rate of the degree of persistence.

\newpage 

\subsection{Single regressor}

Equivalently, to simplify the asymptotics of the following Proposition we consider the univariate predictive regression with a single mildly integrated regressor (no intercept)
\begin{align}
y_{t+1} &=  \beta_1 x_{t} I_{1t} + \beta_2 x_{t} I_{2t} + u_{t+1} 
\\
x_{t} &= \left( 1 - \frac{c}{T^{\gamma}}  \right) x_{t-1} + v_t 
\end{align}
We use the following asymptotic terms
\begin{align}
\frac{ \sum_{t=1}^T x_t^2  }{ T^{1 + \gamma} } &\overset{ p }{  \to } \frac{ \omega^2_v}{2c}
\\
\nonumber
\\
\frac{ \sum_{t=1}^T x_{t} u_{t+1}  }{ T^{ \frac{1 + \gamma}{2}}}  &\overset{ p }{  \to } \mathcal{N} \left( 0,  \frac{ \sigma_u^2 \omega_v^2 }{2c} \right)
\\
\nonumber
\\
\frac{ \sum_{t=1}^T x_{t} I_{1t}  }{ T^{1 + \gamma} }  &\overset{ p }{  \to } \pi \frac{ \omega^2_v}{2c}
\\
\nonumber
\\
\frac{ \sum_{t=1}^T x_{t} I_{1t}  }{ T^{1 + \gamma} }  &\overset{ p }{  \to } (1 - \pi) \frac{ \omega^2_v}{2c}
\end{align}

\begin{proposition}
\label{Proposition1}
Under conditions A1 and A2 of Assumption 1 of the paper, the standard Wald OLS statistic given by the following expression
\begin{align}
\mathcal{W}_T( \pi ) 
&= \frac{1}{ \hat{\sigma}_u^2 } \left( \hat{\beta}_1 -  \hat{\beta}_2  \right)^{\prime}  \left[ \mathcal{R} \left( X^{\prime} X \right)^{-1} \mathcal{R}^{\prime}\right]^{-1} \left( \hat{\beta}_1 - \hat{\beta}_2  \right) 
\end{align}
for testing the null hypothesis $\mathbb{H}_0: \beta_1 = \beta_2$, when the regressor is assumed to be generated via the following process
\begin{align}
x_t = \left( 1 - \frac{c}{ T^{\gamma} } \right) x_{t-1} + v_{t}, \ \ \ \ \text{with} \ \ x_0 = 0, \ \ \text{and} \ \ \gamma \in (0,1). 
\end{align}
is found to have the following limiting distribution
\begin{align}
\mathcal{W}^{*}_T( \pi ) = \underset{ \pi \in [ \pi_1 , \pi_2 ] }{ \text{ sup } } \ \mathcal{W}_T( \pi )  \Rightarrow \displaystyle \underset{ \pi \in [ \pi_1 , \pi_2 ] }{ \text{ sup } } \ \frac{ \bigg[ W( \pi) - \pi W( 1 ) \bigg]^2  }{\pi (1 - \pi)}
\end{align}
\end{proposition}
\begin{remark}
Proposition 1 above shows that when the regressor has persistence properties assumed to be fall in the realm of mildly integrated processes, then the limiting distribution when testing for a structural change in a univariate predictive regression with a single regressor and no intercept, follows a standard NBB limit similar to the classical linear regression case as proved by Andrews (1993). 
\end{remark}

\newpage 

\begin{proof} 
 
We obtain the standard OLS estimators $\hat{\beta}_1$ and $\hat{\beta}_2$ of the corresponding regression coefficients $\beta_1$ and $\beta_2$ as below
\begin{align}
\hat{\beta}_1 &=  \frac{ \displaystyle \sum_{t=1}^T x_{t} I_{1t} y_{t+1} }{ \displaystyle \sum_{t=1}^T x^2_{t} I_{1t} } = \beta^0 +  \frac{ \displaystyle \sum_{t=1}^T x_{t} I_{1t} u_{t+1} }{ \displaystyle \sum_{t=1}^T x^2_{t} I_{1t} } \displaystyle
\\
\hat{\beta}_2 &= \frac{ \displaystyle \sum_{t=1}^T x_{t} I_{2t} y_{t+1} }{ \displaystyle \sum_{t=1}^T x^2_{t} I_{2t} } = \beta^0  +  \frac{ \displaystyle \sum_{t=1}^T x_{t} I_{2t} u_{t+1} }{ \displaystyle \sum_{t=1}^T x^2_{t} I_{2t} }
\end{align} 
Thus, assuming that the structural break is at an unknown break point such as $k = [T \pi]$ for $\pi \in (0,1)$ we consider the limiting results  under the null hypothesis, $\mathbb{H}_0: \beta_1 = \beta_2$. Note that the FCLT does not apply in this case (mildly integrated predictors). However, we use the limit theory already derived in PM. 
\begin{align}
\label{expression1}
T^{ \frac{1 + \gamma}{2}} \left( \hat{\beta}_1 - \beta^0 \right) =  \frac{ \displaystyle \frac{1}{ T^{ \frac{1 + \gamma}{2}} } \sum_{t=1}^{ \floor{ T \pi} } x_{t} u_{t+1} }{ \displaystyle \frac{1}{T^{ 1 + \gamma } } \sum_{t=1}^{ \floor{ T \pi} } x^2_{t} } 
\Rightarrow \frac{ \displaystyle  \mathcal{N} \left( 0,  \pi \frac{ \sigma_u^2 \omega_v^2 }{2c} \right) }{ \displaystyle \pi \frac{ \omega^2_v}{2c} }
\end{align}
\begin{align}
\label{expression2}
T^{ \frac{1 + \gamma}{2}}  \left( \hat{\beta}_2 - \beta^0 \right) =  \frac{ \displaystyle \frac{1}{ T^{ \frac{1 + \gamma}{2}} } \sum_{t=\floor{ T \pi} + 1}^{ T } x_{t} u_{t+1} }{ \displaystyle \frac{1}{T^{ 1 + \gamma } } \sum_{t=\floor{ T \pi} + 1}^{ T } x^2_{t} } 
\Rightarrow \frac{ \displaystyle  \mathcal{N} \left( 0,  (1 - \pi) \frac{ \sigma_u^2 \omega_v^2 }{2c} \right) }{ \displaystyle (1 - \pi) \frac{ \omega^2_v}{2c} }
\end{align}

Thus, using \eqref{expression1} and \eqref{expression2} we obtain the following simplified expression
\begin{align}
T^{ \frac{1 + \gamma}{2}} \left( \hat{\beta}_1 - \hat{\beta}_2 \right) 
\Rightarrow \frac{ \displaystyle  \mathcal{N} \left( 0,  \pi \frac{ \sigma_u^2 \omega_v^2 }{2c} \right) }{ \displaystyle \pi \frac{ \omega^2_v}{2c} } - \frac{ \displaystyle  \mathcal{N} \left( 0,  (1 - \pi) \frac{ \sigma_u^2 \omega_v^2 }{2c} \right) }{ \displaystyle (1 - \pi) \frac{ \omega^2_v}{2c} }
\end{align}

Denoting with $X = [ x_{t} I_{1t} \ \ x_{t} I_{2t} ] \equiv [ X_1 \ X_2 ]$ then the Wald test has an equivalent representation as below
\begin{align}
\mathcal{W}_T( \pi ) 
&= \frac{1}{ \hat{\sigma}_u^2 } \left( \hat{\beta}_1 - \hat{\beta}_2  \right)^{\prime}  \left[ \mathcal{R} \left( X^{\prime} X \right)^{-1} \mathcal{R}^{\prime}\right]^{-1} \left( \hat{\beta}_1 - \hat{\beta}_2 \right) 
\end{align}
Firstly, it can be easily proved that the following equivalent expression holds, using the orthogonality property of $X_1$ and $X_2$ and via a standard matrix inversion application.
\begin{align*}
\left[ \mathcal{R} \left( X^{\prime} X \right)^{-1} \mathcal{R}^{\prime}\right] &= \left[ \left( \sum_{t=1}^T x^2_{t} I_{1t} \right)^{-1} +  \left( \sum_{t=1}^T x^2_{t} I_{1t} \right)^{-1} \right] \\
&= \frac{ \displaystyle \sum_{t=1}^T x^2_{t} I_{1t}  +  \sum_{t=1}^T x^2_{t} I_{2t} }{ \displaystyle \left( \sum_{t=1}^T x^2_{t} I_{1t} \right) \left( \sum_{t=1}^T x_{t} I_{2t} \right) } 
= \frac{ \displaystyle \sum_{t=1}^T x^2_{t} }{ \displaystyle \left( \sum_{t=1}^T x^2_{t} I_{1t} \right) \left( \sum_{t=1}^T x^2_{t} I_{2t} \right) }
\end{align*} 
Therefore, the simplified expression of the Wald statistic is given by the expression below in the case of single predictors 
\begin{align*}
\mathcal{W}_T( \pi ) = \frac{ \left( \hat{\beta}_1 - \hat{\beta}_2 \right)^2 }{ \hat{\sigma}^2_u } \left[  \frac{ \displaystyle \sum_{t=1}^T x^2_{t} }{ \displaystyle \left( \sum_{t=1}^T x^2_{t} I_{1t} \right) \left( \sum_{t=1}^T x^2_{t} I_{2t} \right) } \right]^{-1} 
&= 
\frac{ \left( \hat{\beta}_1 - \hat{\beta}_2 \right)^2 }{ \hat{\sigma}^2_u } \frac{  \displaystyle \left( \sum_{t=1}^T x^2_{t} I_{1t} \right) \left( \sum_{t=1}^T x^2_{t} I_{2t} \right) }{  \displaystyle \sum_{t=1}^T x^2_{t} }   \\
&=  
T^{ 1 + \gamma } \frac{ \left( \hat{\beta}_1 - \hat{\beta}_2 \right)^2 }{ \hat{\sigma}^2_u }  \frac{  \displaystyle \left( \sum_{t=1}^T \frac{ x^2_{t} I_{1t}}{ T^{ 1 + \gamma } } \right) \left( \sum_{t=1}^T \frac{ x^2_{t} I_{2t} }{ T^{ 1 + \gamma } } \right)  }{  \displaystyle \sum_{t=1}^T   \frac{ x^2_{t} }{ T^{ 1 + \gamma } } } 
\end{align*}
Moreover, the following asymptotic convergence result also holds
\begin{align}
\frac{ \displaystyle \left( \sum_{t=1}^T \frac{ x^2_{t} I_{1t}}{ T^{ 1 + \gamma } } \right) \left( \sum_{t=1}^T \frac{ x^2_{t} I_{2t} }{ T^{ 1 + \gamma } } \right)  }{ \displaystyle \sum_{t=1}^T   \frac{ x^2_{t} }{ T^{ 1 + \gamma } } } \Rightarrow \frac{ \displaystyle  \pi \frac{ \omega^2_v}{2c} (1 - \pi) \frac{ \omega^2_v}{2c} }{ \displaystyle \frac{ \omega^2_v}{2c} } = \pi(1 - \pi) \frac{ \omega^2_v}{2c}
\end{align}
Thus, we obtain that
\begin{align}
\mathcal{W}_T( \pi ) 
&\equiv \frac{1}{ \sigma^2_u } \left\{ \frac{ \displaystyle  \mathcal{N} \left( 0,  \pi \frac{ \sigma_u^2 \omega_v^2 }{2c} \right) }{ \displaystyle \pi \frac{ \omega^2_v}{2c} } - \frac{ \displaystyle  \mathcal{N} \left( 0,  (1 - \pi) \frac{ \sigma_u^2 \omega_v^2 }{2c} \right) }{ \displaystyle (1 - \pi) \frac{ \omega^2_v}{2c} }  \right\}^2 \pi(1 - \pi) \frac{ \omega^2_v}{2c}
\nonumber
\\
\nonumber
\\
&= 
\frac{1}{ \sigma^2_u } \frac{ \displaystyle \left\{ (1 - \pi) \ \mathcal{N} \left( 0,  \pi \frac{ \sigma_u^2 \omega_v^2 }{2c} \right) - \pi \ \mathcal{N} \left( 0,  (1 - \pi) \frac{ \sigma_u^2 \omega_v^2 }{2c} \right) \right\}^2 }{ \displaystyle \pi (1 - \pi) \frac{ \omega^2_v}{2c} }   
\end{align}

\newpage 

Thus, simplifying the terms which do not depend on $\pi$ from above expression and using the supremum functional as well since we consider an unknown break-point then we obtain the following expression
\begin{align*}
\mathcal{W}^{*}_T( \pi ) 
&\Rightarrow \underset{ \pi \in [ \pi_1 , \pi_2 ] }{ \text{ sup } } \  \frac{ \displaystyle \left\{ (1 - \pi) \ \mathcal{N} \bigg( 0,  \pi  \bigg) - \pi \ \mathcal{N} \bigg( 0,  (1 - \pi) \bigg) \right\}^2 }{ \displaystyle \pi (1 - \pi) }   
\nonumber
\\
\nonumber
\\
\nonumber
&\equiv 
\underset{ \pi \in [ \pi_1 , \pi_2 ] }{ \text{ sup } } \  \frac{ \displaystyle \left\{ \mathcal{N} \bigg( 0,  \pi  \bigg) - \pi \ \mathcal{N} \bigg( 0, 1 \bigg) \right\}^2 }{ \displaystyle \pi (1 - \pi) }
\end{align*}
which shows indeed the weakly convergence to a NBB in the case of a single mildly integrated predictor in the predictive regression model. 
\end{proof}

In summary, we show that the sup Wald-OLS statistic for testing for a single structural change predictive regressions with mildly integrated predictors, which is the case that the degree of persistence is controlled via the exponent rate $\gamma \in (0,1)$, then we obtain an asymptotically equivalent limiting distribution as in the standard linear regression.

Furthermore, since $\gamma \in (0,1)$ and assuming that the exponent rate for the degree of persistence of the instrument satisfies $\gamma \in ( 0 , \delta )$, then using Lemma 3.5 of PM the following asymptotic terms hold
\begin{align}
\frac{1}{ T^{1 + \gamma}  } \sum_{t=1}^T x_t \tilde{z}_t^{\prime} 
&= \frac{1}{ T^{1 + \gamma}  } \sum_{t=1}^T x_t  x_t^{\prime} + o_p(1)
\\
\frac{1}{ T^{1 + \gamma}  } \sum_{t=1}^T \tilde{z}_t  \tilde{z}_t^{\prime} &=  \frac{1}{ T^{1 + \gamma}  } \sum_{t=1}^T x_t  x_t^{\prime} + o_p(1)
\end{align}

\subsection{Multiple regressors}

Next, we consider the case of multiple mildly integrated regressors. First, we consider the following example, which provides useful insights for the related asymptotic terms in the case of the univariate predictive regression with multiple predictors.
\begin{example}
\label{example1.appendix}
Consider the predictive regression with multiple predictors given below
\begin{align}
y_{t+1} = \beta_0 + \underline{\beta}_1^{\prime} \ \underline{x}_t + u_{t+1}
\end{align}
We aim to examine the limiting distribution of the parameter vector $\underline{\beta} = ( \beta_0, \underline{\beta}_1^{\prime} )$ with $\beta_0 \in \mathbb{R}$, $\underline{\beta}_1 \in \mathbb{R}^{p \times 1} $ and $\widetilde{x}_t = \left( \underline{1} , \underline{x}^{\prime}_t \right)^{\prime} \in \mathbb{R}^{ T \times (p+1) }$. Note that since the intercept and the vector of predictors have a different convergence rate, we define the normalization matrix: $\mathcal{D}_T = diag( \sqrt{T}, T^{ \frac{ 1 + \gamma }{2} } \text{I}_{p} )\in \mathbb{R}^{(p+1) \times (p+1)}$. 

\newpage 

We have that
\begin{align}
\left( \underline{ \widehat{\beta} } - \underline{\beta}^0 \right) = \left(  \sum_{t=1}^T \widetilde{ \underline{x} }_{t} \widetilde{ \underline{x} }_{t}^{\prime} \right)^{-1}  \left( \sum_{t=1}^T \widetilde{ \underline{x} }_{t} u_{t+1} \right) 
\end{align}
where $\underline{\beta}^0$ the true parameter vector under the null hypothesis, $\mathbb{H}_0: \underline{\beta} = \underline{\beta}^0$. 
We obtain
\begin{align}
\mathcal{D}_T^{-1} \left( \sum_{t=1}^T \widetilde{\underline{x}}_{t} \widetilde{\underline{x}}_{t}^{\prime} \right) \mathcal{D}_T^{-1} 
=
\begin{bmatrix}
1 &  \displaystyle  \frac{1}{ T^{ \frac{ 1 + \gamma }{2} } } \sum_{t=1}^T \underline{x}_{t}^{\prime} 
\\
\\
\displaystyle \frac{1}{ T^{ \frac{ 1 + \gamma }{2} } } \sum_{t=1}^T  \underline{x}_{t} & \displaystyle \frac{1}{ T^{ 1 + \gamma }} \sum_{t=1}^T  \underline{x}_{t} \underline{x}_{t}^{\prime}
\end{bmatrix}
\Rightarrow
\begin{bmatrix}
1 &  \displaystyle \textcolor{red}{0}
\\
\\
\displaystyle \textcolor{red}{0} &  \displaystyle V_{zz}
\end{bmatrix}
\end{align}
Similarly, we also obtain the following expression
\begin{align}
\mathcal{D}_T^{-1} \left( \sum_{t=1}^T \widetilde{ \underline{x} }_{t} u_{t+1} \right) 
=
\begin{bmatrix}
\displaystyle \frac{1}{ \sqrt{T} } \sum_{t=1}^T u_{t+1}
\\
\displaystyle \frac{1}{ T^{ \frac{ 1 + \gamma }{2} } } \sum_{t=1}^T \underline{x}^{\prime}_t u_{t+1}
\end{bmatrix} 
\Rightarrow 
\begin{bmatrix}
B_u (r) 
\\
\\
\displaystyle \mathcal{N} \bigg( 0, \sigma_u^2 V_{zz} \bigg)
\end{bmatrix}
\end{align}
Therefore, combining the above results, we obtain 
\begin{align}
\mathcal{D}_T \left( \underline{ \widehat{\beta} } - \underline{\beta}^0 \right) 
&= \left[ \mathcal{D}_T^{-1} \left(  \sum_{t=1}^T \widetilde{ \underline{x} }_{t} \widetilde{ \underline{x} }_{t}^{\prime} \right) \mathcal{D}_T^{-1} \right]^{-1} \mathcal{D}_T^{-1} \left( \sum_{t=1}^T \widetilde{ \underline{x} }_{t}^{\prime} u_{t+1} \right) 
\nonumber
\\
&\Rightarrow
\begin{bmatrix}
1 &  \displaystyle \textcolor{red}{0}
\\
\\
\displaystyle \textcolor{red}{0} &  \displaystyle V_{zz}
\end{bmatrix}^{-1} 
\times
\begin{bmatrix}
B_u (r) 
\\
\\
\displaystyle \mathcal{N} \bigg( 0, \sigma_u^2 V_{zz} \bigg)
\end{bmatrix}
\end{align}
\end{example}
\begin{proposition}
\label{Proposition2}
Under conditions A1 and A2 of Assumption 1 of the paper, the standard Wald OLS statistic given by the following expression
\begin{align}
\mathcal{W}_T( \pi ) 
&= \frac{1}{ \hat{\sigma}_u^2 } \left( \underline{ \widehat{\beta} }_1 -  \underline{ \widehat{\beta} }_2  \right)^{\prime}  \left[ \mathcal{R} \left( X^{\prime} X \right)^{-1} \mathcal{R}^{\prime}\right]^{-1} \left( \underline{ \widehat{\beta} }_1 -  \underline{ \widehat{\beta} }_2  \right) 
\end{align}
for testing the null hypothesis $\mathbb{H}_0: \underline{\beta}_1 = \underline{\beta}_2$, when the regressor is generated via 
\begin{align}
\underline{x}_t = \left( I_p - \frac{C}{ T^{\gamma} } \right) \underline{x}_{t-1} + \underline{v}_{t}, \ \ \ \ \text{with} \ \ \underline{x}_0 = 0, \ \ \text{and} \ \ \gamma \in (0,1). 
\end{align}
is found to have the following limiting distribution
\begin{align}
\mathcal{W}_T( \pi )  \Rightarrow  \chi^2_k +  \underset{ \pi \in [ \pi_1 , \pi_2 ] }{ \text{ sup } } \ \ \frac{ \displaystyle \mathcal{B B} (\pi)  }{ \pi (1 - \pi)   }
\end{align}
\end{proposition}

\newpage 

\begin{proof}

Under the null hypothesis of no structural break, $\mathbb{H}_0: \underline{\beta}_1 = \underline{\beta}_2$, we have 
\begin{align*}
\mathcal{D}_T \left( \underline{ \widehat{\beta} }_1 - \underline{\beta}^0 \right) 
&= \left[ \mathcal{D}_T^{-1} \left(  \sum_{t=1}^T \widetilde{ \underline{x} }_{t} \widetilde{ \underline{x} }_{t}^{\prime} I_{1t} \right) \mathcal{D}_T^{-1} \right]^{-1} \mathcal{D}_T^{-1} \left( \sum_{t=1}^T \widetilde{ \underline{x} }_{t}^{\prime} u_{t+1} I_{1t} \right)
\end{align*}
where $\underline{\beta}^0$, the population value of the model coefficient.  

Similarly, 
\begin{align*}
\mathcal{D}_T \left( \underline{ \widehat{\beta} }_2 - \underline{\beta}^0 \right) 
&= \left[ \mathcal{D}_T^{-1} \left(  \sum_{t=1}^T \widetilde{ \underline{x} }_{t} \widetilde{ \underline{x} }_{t}^{\prime} I_{2t} \right) \mathcal{D}_T^{-1} \right]^{-1} \mathcal{D}_T^{-1} \left( \sum_{t=1}^T \widetilde{ \underline{x} }_{t}^{\prime} u_{t+1} I_{2t} \right)
\end{align*}

The weakly convergence result for the estimator of $\beta_1$ follows
\begin{align}
\label{expression6}
\mathcal{D}_T \left( \underline{ \widehat{\beta} }_1 - \underline{\beta}^0 \right) 
&\Rightarrow 
\begin{bmatrix}
1 &  \displaystyle \textcolor{red}{0}
\\
\\
\displaystyle \textcolor{red}{0} &  \displaystyle \pi V_{zz}
\end{bmatrix}^{-1} 
\times
\begin{bmatrix}
B_u (\pi) 
\\
\\
\displaystyle \mathcal{N} \bigg( 0, \pi \sigma_u^2 V_{zz} \bigg)
\end{bmatrix} 
\end{align}
Similarly, for the estimator of $\beta_2$ we have the following weakly convergence result
\begin{align}
\label{expression7}
\mathcal{D}_T \left( \underline{ \widehat{\beta} }_2 - \underline{\beta}^0 \right) 
&\Rightarrow 
\begin{bmatrix}
1 &  \displaystyle \textcolor{red}{0}
\\
\\
\displaystyle \textcolor{red}{0} &  \displaystyle (1 - \pi) V_{zz}
\end{bmatrix}^{-1} 
\times
\begin{bmatrix}
B_u (1) - B_u (\pi) 
\\
\\
\displaystyle \mathcal{N} \bigg( 0, (1 - \pi) \sigma_u^2 V_{zz} \bigg)
\end{bmatrix}
\end{align}
Therefore, we have that
\begin{align}
\mathcal{D}_T \left( \underline{ \widehat{\beta} }_1 - \underline{\beta}^0 \right) - \mathcal{D}_T \left( \underline{ \widehat{\beta} }_2 - \underline{\beta}^0 \right)   
\equiv 
\mathcal{D}_T \left( \underline{ \widehat{\beta} }_1 - \underline{ \widehat{\beta} }_2 \right)
\end{align}

Recall that the expression for the Wald statistic is as below
\begin{align}
\mathcal{W}_T( \pi ) 
&= 
\frac{1}{ \hat{\sigma}_u^2 } \left( \underline{ \widehat{\beta} }_1 - \underline{ \widehat{\beta} }_2 \right)^{\prime}  \left[ \mathcal{R} \left( X^{\prime} X \right)^{-1} \mathcal{R}^{\prime}\right]^{-1} \left( \underline{ \widehat{\beta} }_1 - \underline{ \widehat{\beta} }_2 \right)
\end{align}
Note that, the following holds
\begin{align}
\left[ \mathcal{R} \left( X^{\prime} X \right)^{-1} \mathcal{R}^{\prime}\right] 
&=  \left[ \left( X_1^{\prime} X_1 \right)^{-1} + \left( X_2^{\prime} X_2 \right)^{-1} \right]
\nonumber
\\
&=  \mathcal{D}_T^{-1}  \left[ \bigg( \mathcal{D}_T^{-1} \left( X_1^{\prime} X_1 \right) \mathcal{D}_T^{-1} \bigg)^{-1} + \bigg( \mathcal{D}_T^{-1} \left( X_2^{\prime} X_2 \right) \mathcal{D}_T^{-1} \bigg)^{-1} \right] \mathcal{D}_T^{-1} 
\end{align}

\newpage

Thus, the Wald statistic is expressed as below
\begin{align}
\mathcal{W}_T( \pi ) 
&= 
\frac{1}{ \hat{\sigma}_u^2 } \left( \mathcal{D}_T \left( \underline{ \widehat{\beta} }_1 - \underline{ \widehat{\beta} }_2 \right)\right) ^{\prime}  \mathcal{D}_T^{-1}  \left[ \bigg( \mathcal{D}_T^{-1} \left( X_1^{\prime} X_1 \right) \mathcal{D}_T^{-1} \bigg)^{-1} + \bigg( \mathcal{D}_T^{-1} \left( X_2^{\prime} X_2 \right) \mathcal{D}_T^{-1} \bigg)^{-1} \right]^{-1}  \mathcal{D}_T^{-1} 
\nonumber
\\
&\ \ \ \ \ \times \mathcal{D}_T \left( \underline{ \widehat{\beta} }_1 - \underline{ \widehat{\beta} }_2 \right)
\end{align}

Now, we can consider the limiting distribution of the Wald OLS statistic by replacing the asymptotic terms for both the distance measure and the covariance matrix in the expression for the test statistic. We obtain the following 
\begin{align}
\mathcal{D}_T \left( \underline{ \widehat{\beta} }_1 - \underline{ \widehat{\beta} }_2 \right)
&=
\begin{bmatrix}
1 &  \displaystyle \textcolor{red}{0}
\\
\\
\displaystyle \textcolor{red}{0} &  \displaystyle \pi V_{zz}
\end{bmatrix}^{-1} 
\times
\begin{bmatrix}
B_u (\pi) 
\\
\\
\displaystyle \mathcal{N} \bigg( 0, \pi \sigma_u^2 V_{zz} \bigg)
\end{bmatrix}
\nonumber
\\
&- 
\begin{bmatrix}
1 &  \displaystyle \textcolor{red}{0}
\\
\\
\displaystyle \textcolor{red}{0} &  \displaystyle (1 - \pi) V_{zz}
\end{bmatrix}^{-1} 
\times
\begin{bmatrix}
B_u (1) - B_u (\pi) 
\\
\\
\displaystyle \mathcal{N} \bigg( 0, (1 - \pi) \sigma_u^2 V_{zz} \bigg)
\end{bmatrix}
\end{align}

Note that the off-diagonal elements converge in probability to zero. Thus, we obtain
\begin{align}
\mathcal{D}_T \left( \underline{ \widehat{\beta} }_1 - \underline{ \widehat{\beta} }_2 \right)
&=
\begin{bmatrix}
1 & 0
\\
\\
0 &  \displaystyle \frac{1}{ \pi } V_{zz}^{-1}
\end{bmatrix} 
\times
\begin{bmatrix}
B_u (\pi) 
\\
\\
\displaystyle \mathcal{N} \bigg( 0, \pi \sigma_u^2 V_{zz} \bigg)
\end{bmatrix}
\nonumber
\\
&- 
\begin{bmatrix}
1 &  0
\\
\\
0 &  \displaystyle  \frac{1}{1 - \pi} V_{zz}^{-1}
\end{bmatrix} 
\times
\begin{bmatrix}
B_u (1) - B_u (\pi) 
\\
\\
\displaystyle \mathcal{N} \bigg( 0, (1 - \pi) \sigma_u^2 V_{zz} \bigg)
\end{bmatrix}
\nonumber
\\
\nonumber
\\
&= 
\begin{bmatrix}
B_u (\pi) 
\\
\\
\displaystyle \frac{1}{ \pi } V_{zz}^{-1} \mathcal{N} \bigg( 0, \pi \sigma_u^2 V_{zz} \bigg)
\end{bmatrix}
- 
\begin{bmatrix}
B_u (1) - B_u (\pi) 
\\
\\
\displaystyle \frac{1}{1 - \pi } V_{zz}^{-1} \mathcal{N} \bigg( 0, (1 - \pi) \sigma_u^2 V_{zz} \bigg)
\end{bmatrix}
\nonumber
\\
\nonumber
\\
&= 
\begin{bmatrix}
- B_u (1) 
\\
\\
\displaystyle \frac{1}{ \pi } V_{zz}^{-1} \mathcal{N} \bigg( 0, \pi \sigma_u^2 V_{zz} \bigg)  - \frac{1}{1 - \pi } V_{zz}^{-1} \mathcal{N} \bigg( 0, (1 - \pi) \sigma_u^2 V_{zz} \bigg)
\end{bmatrix}
\end{align}

\newpage 

which implies that 
\begin{align}
\mathcal{D}_T \left( \underline{ \widehat{\beta} }_1 - \underline{ \widehat{\beta} }_2 \right)
&=
\begin{bmatrix}
- B_u (1) 
\\
\\
\displaystyle \frac{ V_{zz}^{-1} }{ \pi (1 - \pi) }  \left\{ \mathcal{N} \bigg( 0, \pi \sigma_u^2 V_{zz} \bigg)  -\pi \ \mathcal{N} \bigg( 0,  \sigma_u^2 V_{zz} \bigg) \right\}
\end{bmatrix}
\end{align}

The asymptotic convergence of the covariance matrix is given by \
\begin{align}
\left\{
\begin{bmatrix}
1 & 0
\\
\\
0 &  \displaystyle \frac{1}{ \pi } V_{zz}^{-1}
\end{bmatrix} 
+
\begin{bmatrix}
1 & 0
\\
\\
0 &  \displaystyle \frac{1}{1 - \pi } V_{zz}^{-1}
\end{bmatrix} 
\right\}^{-1}
&= 
\begin{bmatrix}
1 & 0
\\
\\
0 &  \displaystyle \frac{1}{\pi( 1 - \pi) } V_{zz}^{-1}
\end{bmatrix}^{-1}
\nonumber
\\
\nonumber
\\
&= 
\begin{bmatrix}
1 & 0
\\
\\
0 &  \displaystyle \pi( 1 - \pi) V_{zz}
\end{bmatrix}
\end{align}

Therefore, we obtain
\begin{align}
\mathcal{W}_T( \pi ) 
&\Rightarrow
\frac{1}{ \sigma_u^2 }
\begin{bmatrix}
- B_u (1) 
\\
\\
\displaystyle \frac{ V_{zz}^{-1} }{ \pi (1 - \pi) }  \left\{ \mathcal{N} \bigg( 0, \pi \sigma_u^2 V_{zz} \bigg)  -\pi \ \mathcal{N} \bigg( 0,  \sigma_u^2 V_{zz} \bigg) \right\}
\end{bmatrix}^{\prime}
\begin{bmatrix}
1 & 0
\\
\\
0 &  \displaystyle \pi( 1 - \pi) V_{zz}
\end{bmatrix}
\nonumber
\\
\nonumber
\\
&\ \ \ \ \ \ \times
\begin{bmatrix}
- B_u (1) 
\\
\\
\displaystyle \frac{ V_{zz}^{-1} }{ \pi (1 - \pi) }  \left\{ \mathcal{N} \bigg( 0, \pi \sigma_u^2 V_{zz} \bigg)  -\pi \ \mathcal{N} \bigg( 0,  \sigma_u^2 V_{zz} \bigg) \right\}
\end{bmatrix}
\nonumber
\\
\nonumber
\\
&= 
\begin{bmatrix}
- W_u (1) , 
 & 
\displaystyle V_{zz}^{-1} \left\{ \mathcal{N} \bigg( 0, \pi V_{zz} \bigg)  -\pi \ \mathcal{N} \bigg( 0,  V_{zz} \bigg) \right\}
\end{bmatrix}
\begin{bmatrix}
- W_u (1) 
\\
\displaystyle \frac{ V_{zz}^{-1} }{ \pi (1 - \pi) }  \left\{ \mathcal{N} \bigg( 0, \pi V_{zz} \bigg)  -\pi \ \mathcal{N} \bigg( 0, V_{zz} \bigg) \right\}
\end{bmatrix}
\nonumber
\\
\nonumber
\\
&= \bigg( W_u (1) \bigg)^2 +  \underset{ \pi \in [ \pi_1 , \pi_2 ] }{ \text{ sup } } \  \frac{ \displaystyle \big[ W(\pi) - \pi W(1) \big]^{\prime} \big[ W(\pi) - \pi W(1) \big]  }{ \pi (1 - \pi)   }
\nonumber
\\
\nonumber
\\
&:= \chi^2_p +  \underset{ \pi \in [ \pi_1 , \pi_2 ] }{ \text{ sup } } \ \ \frac{ \displaystyle \mathcal{B B} (\pi)  }{ \pi (1 - \pi)   }
\end{align}

\end{proof}


\newpage 

\section{Asymptotic Distribution of sup Wald-IVX statistic}

Consider the univariate predictive regression with multiple predictors \begin{align}
y_{t+1} = \left( \alpha_1 + \beta_1 x_{t} \right) I_{1t} + \left( \alpha_2 + \beta_2 x_{t} \right) I_{2t} + u_{t+1}
\end{align}
and $x_t$ is generated via a LUR process as below
\begin{align}
x_t = \left(I_p - \frac{C}{T} \right) x_{t-1} + v_{t}, \ \ \ \ \text{with} \ \ x_0 = 0. 
\end{align}
where $I_{1t} :=  \mathbf{1} \{ t \leq k  \}$ and $I_{2t} :=  \mathbf{1} \{ t > k  \}$ with $k = \floor{ T\pi}$.

Denote with $X_1 \in \mathbb{R}^{T \times (p+1)}$ to represent the corresponding matrix stacking $I_{1t}$, i.e., $x_t I_{1t} \equiv x_{1t}$ and similarly $X_2 \in \mathbb{R}^{T \times (p+1)}$ represents the matrix stacking $I_{2t}$, i.e, $x_t I_{2t} \equiv x_{2t}$ including a column with one's to capture the model intercept in both regimes. We consider the following two normalization matrices, that is, $\mathcal{D}_1 = \text{diag}\left(  T^{ \frac{\delta}{2} },  T^{ \frac{ 1 + \delta }{ 2 } } \text{I}_p \right)$ and $\mathcal{D}_2 = \text{diag}\left(  T^{ 1 - \frac{\delta}{2} },  T^{ \frac{ 1 + \delta }{ 2 } } \text{I}_p \right)$, where $\text{I}_p$ the $( p \times p)$ identity matrix. 

The IVX instrumentation implies that
\begin{align}
\tilde{ z }_t = \sum_{j=1}^t R_{Tz}^{t-j} \Delta x_j, \ \ R_{Tz} = \left( \text{I}_p - \frac{C_z}{ T^{\delta} } \right), \delta \in (0,1), C_z > 0.
\end{align}     

Note that all matrices, $\left\{ X_1, X_2, Z_1, Z_2 \right\} \in \mathbb{R}^{ T \times (p+1)}$, include the first column to be a column vector of ones with the remaining columns to represent the corresponding stacked values from the set of $p-$predictors which are included in the model. Furthermore, for simplicity of notation, we consider that $Z_j$ for $j = 1,2$ represents the corresponding IVX instruments, constructed via the IVX instrumentation procedure of PM.  Moreover, we operate under the assumption of an unknown break-point $\pi \in \Pi$.

The Wald IVX statistic has the following form 
\begin{align*}
\mathcal{W}_T( \pi) 
= \frac{1}{ \hat{\sigma}^2 }
 \left[ \left( {Z}_1^{\prime} X_1 \right)^{-1} {Z}_1^{\prime} u  - \left( {Z}_2^{\prime} X_2 \right)^{-1} {Z}_2^{\prime} u \right]^{\prime} \mathcal{Q}_{ \mathcal{R} }^{-1} \left[ \left( {Z}_1^{\prime} X_1 \right)^{-1} {Z}_1^{\prime} u  - \left( {Z}_2^{\prime} X_2 \right)^{-1} {Z}_2^{\prime} u \right]
\end{align*}
where the  covariance matrix $\mathcal{Q}_{ \mathcal{R} }$ is defined as below
\begin{align}
\label{Qmatrix}
\mathcal{Q}_{ \mathcal{R} } := \left\{  \left(Z_1^{\prime} X_1 \right)^{-1} \left( Z_1^{\prime} Z_1 \right) \left(X_1^{\prime} Z_1 \right)^{-1}  +  \left(Z_2^{\prime} X_2 \right)^{-1} \left( Z_2^{\prime} Z_2 \right) \left( X_2^{\prime} Z_2 \right)^{-1}  \right\} 
\end{align}
We denote with $\theta_i = ( \alpha_i, \beta_i )$ for $i = 1,2$, the IVX estimator which is expressed as below
\begin{align*}
\left( \hat{\theta}_1 - \theta  \right) 
&= \left( {Z}_1^{\prime} X_1 \right)^{-1} {Z}_1^{\prime} u 
\\
&\equiv \mathcal{D}_1^{-1} \left[ \mathcal{D}_2^{-1} \left( {Z}_1^{\prime} X_1 \right) \mathcal{D}_1^{-1} \right]^{-1} \mathcal{D}_2^{-1} \left( {Z}_1^{\prime} u \right)
\end{align*}
Thus, 
\begin{align*}
\mathcal{D}_1 \left( \hat{\theta}_1 - \theta  \right) = \left[ \mathcal{D}_2^{-1} \left( {Z}_1^{\prime} X_1 \right) \mathcal{D}_1^{-1} \right]^{-1} \mathcal{D}_2^{-1} \left( {Z}_1^{\prime} u \right)
\end{align*}
Similarly, 
\begin{align*}
\left( \hat{\theta}_2 - \theta  \right) 
&= \left( {Z}_2^{\prime} X_2 \right) {Z}_2^{\prime} u 
\\
&\equiv \mathcal{D}_1^{-1} \left[ \mathcal{D}_2^{-1} \left( {Z}_2^{\prime} X_2 \right) \mathcal{D}_1^{-1} \right]^{-1} \mathcal{D}_2^{-1} \left( {Z}_2^{\prime} u \right)
\end{align*}
Thus, 
\begin{align*}
\mathcal{D}_1 \left( \hat{\theta}_2 - \theta  \right) = \left[ \mathcal{D}_2^{-1} \left( {Z}_2^{\prime} X_2 \right) \mathcal{D}_1^{-1} \right]^{-1} \mathcal{D}_2^{-1} \left( {Z}_2^{\prime} u \right)
\end{align*}
Therefore, 
\begin{align*}
\mathcal{D}_1 \left( \hat{\theta}_1 - \hat{\theta}_2  \right) = \left\{ \left[ \mathcal{D}_2^{-1} \left( {Z}_1^{\prime} X_1 \right) \mathcal{D}_1^{-1} \right]^{-1} \mathcal{D}_2^{-1} \left( {Z}_1^{\prime} u \right) - \left[ \mathcal{D}_2^{-1} \left( {Z}_2^{\prime} X_2 \right) \mathcal{D}_1^{-1} \right]^{-1} \mathcal{D}_2^{-1} \left( {Z}_2^{\prime} u \right) \right\}
\end{align*}
and/or equivalently,
\begin{align*}
\left( \hat{\theta}_1 - \hat{\theta}_2  \right) = \left\{ \mathcal{D}_1^{-1} \left[ \mathcal{D}_2^{-1} \left( {Z}_1^{\prime} X_1 \right) \mathcal{D}_1^{-1} \right]^{-1} \mathcal{D}_2^{-1} \left( {Z}_1^{\prime} u \right) - \mathcal{D}_1^{-1} \left[ \mathcal{D}_2^{-1} \left( {Z}_2^{\prime} X_2 \right) \mathcal{D}_1^{-1} \right]^{-1} \mathcal{D}_2^{-1} \left( {Z}_2^{\prime} u \right) \right\}
\end{align*}
Furthermore, for the covariance matrix we have that
\begin{align*}
\mathcal{D}_1 \mathcal{Q}_{ \mathcal{R} } \mathcal{D}_2 
&=
\mathcal{D}_1 \left(Z_1^{\prime} X_1 \right)^{-1} \left( Z_1^{\prime} Z_1 \right) \left(X_1^{\prime} Z_1 \right)^{-1} \mathcal{D}_2  +  \mathcal{D}_1 \left(Z_2^{\prime} X_2 \right)^{-1} \left( Z_2^{\prime} Z_2 \right) \left( X_2^{\prime} Z_2 \right)^{-1}  \mathcal{D}_2
\\
&\equiv
\bigg[ \mathcal{D}_2^{-1} \left(Z_1^{\prime} X_1 \right) \mathcal{D}_1^{-1} \bigg]^{-1}  \bigg[ \mathcal{D}_2^{-1} \left( Z_1^{\prime} Z_1 \right) \mathcal{D}_1^{-1} \bigg] \bigg[ \mathcal{D}_2^{-1} \left(Z_1^{\prime} X_1 \right) \mathcal{D}_1^{-1} \bigg]^{-1}
\nonumber
\\
&\ +  
\bigg[ \mathcal{D}_2^{-1} \left(Z_2^{\prime} X_2 \right) \mathcal{D}_1^{-1} \bigg]^{-1}  \bigg[ \mathcal{D}_2^{-1} \left( Z_2^{\prime} Z_2 \right) \mathcal{D}_1^{-1} \bigg] \bigg[ \mathcal{D}_2^{-1} \left(Z_2^{\prime} X_2 \right) \mathcal{D}_1^{-1} \bigg]^{-1} 
\end{align*}


Therefore, 
\begin{align*}
\mathcal{D}_2 \left( \hat{\theta}_1 - \hat{\theta}_2  \right) = \left\{ \left[ \mathcal{D}_1^{-1} \left( {Z}_1^{\prime} X_1 \right) \mathcal{D}_2^{-1} \right]^{-1} \mathcal{D}_1^{-1} \left( {Z}_1^{\prime} u \right) - \left[ \mathcal{D}_1^{-1} \left( {Z}_2^{\prime} X_2 \right) \mathcal{D}_2^{-1} \right]^{-1} \mathcal{D}_1^{-1} \left( {Z}_2^{\prime} u \right) \right\}
\end{align*}
which gives
\begin{align}
\mathcal{W}^{\text{IVX}}_T( \pi) 
= 
\frac{1}{ \hat{\sigma}^2 } \left( \hat{\theta}_1 - \hat{\theta}_2 \right)^{\prime} \mathcal{D}_2 \bigg[ \mathcal{D}_1  \mathcal{Q}_{ \mathcal{R} }   \mathcal{D}_2 \bigg]^{-1} \mathcal{D}_1 \left( \hat{\theta}_1 - \hat{\theta}_2 \right)
\end{align}

\color{black}

\newpage 

\subsection{Wald IVX test for mildly integrated regressors}

Consider the univariate predictive regression with multiple predictors \begin{align*}
y_{t+1} = \left( \alpha_1 + \beta_1 x_{t} \right) I_{1t} + \left( \alpha_2 + \beta_2 x_{t} \right) I_{2t} + u_{t+1}
\end{align*}
and $x_t$ is generated via the following process
\begin{align*}
x_t = \left(I_p - \frac{C}{ T^{\gamma} } \right) x_{t-1} + v_{t}, \ \ \ \ \text{with} \ \ x_0 = 0. 
\end{align*}
where $I_{1t} :=  \mathbf{1} \{ t \leq k  \}$ and $I_{2t} :=  \mathbf{1} \{ t > k  \}$ with $k = \floor{ T\pi}$.

\textit{Proof.} We consider the weakly convergence of the following sample moments
\begin{align}
\mathcal{D}_2^{-1} \left( Z_1^{\prime} X_1 \right) \mathcal{D}_1^{-1} 
&= 
\begin{bmatrix}
k / T &  \displaystyle  \frac{1}{ T^{ 3 / 2 } } \sum_{t=1}^T  x^{\prime}_{1t}   
\\
\nonumber
\\
\displaystyle \displaystyle \frac{1}{ T^{ \frac{1}{2} + \delta}} \sum_{t=1}^T z_{1t}   & \displaystyle \frac{1}{ T^{1 + \delta } } \sum_{t=1}^T z_{1t} x_{1t}^{\prime}
\end{bmatrix}
\Rightarrow
\begin{bmatrix}
\pi_0 & \ \ \displaystyle  0
\\
\\
\underline{J}_c( \pi_0 ) & \ \ \displaystyle - \pi_0 \Omega_{vv}
\end{bmatrix} 
\\
\\
\mathcal{D}_2^{-1} \left( Z_2^{\prime} X_2 \right) \mathcal{D}_1^{-1} 
&= 
\begin{bmatrix}
1 - k / T &  \displaystyle  \frac{1}{ T^{ 3 / 2 } } \sum_{t=1}^T  x^{\prime}_{2t}   
\\
\\
\displaystyle \displaystyle \frac{1}{ T^{ \frac{1}{2} + \delta}} \sum_{t=1}^T z_{2t}   & \displaystyle \frac{1}{ T^{1 + \delta } } \sum_{t=1}^T z_{2t} x_{2t}^{\prime}
\end{bmatrix}
\Rightarrow
\begin{bmatrix}
(1 - \pi_0) & \ \ \displaystyle 0
\\
\\
\underline{J}_c( 1 ) - \underline{J}_c( \pi_0 )  & \ \  \displaystyle - (1 - \pi_0) \Omega_{vv}
\end{bmatrix} 
\end{align}
\begin{align}
\mathcal{D}_2^{-1} \left( \sum_{t=1}^T Z_{1}^{\prime} Z_1 \right) \mathcal{D}_1^{-1} = \begin{bmatrix}
k / T &  \displaystyle \frac{1}{ T^{3 / 2 } } \sum_{t=1}^T z_{1t}^{\prime}
\\
\\
\displaystyle \displaystyle \frac{1}{ T^{ \frac{1}{2} + \delta}} \sum_{t=1}^T z_{1t} & \displaystyle \frac{1}{ T^{1 + \delta } } \sum_{t=1}^T z_{1t} z_{1t}^{\prime}
\end{bmatrix}
\Rightarrow
\begin{bmatrix}
\pi_0 &  \ \ 0
\\
\\
\underline{J}_{c} ( \pi_0 )   &  \pi_0 V_{zz}  
\end{bmatrix}
\end{align}
\begin{align}
\mathcal{D}_2^{-1} \left( \sum_{t=1}^T Z_{2}^{\prime} Z_2 \right) \mathcal{D}_1^{-1} 
= \begin{bmatrix}
1 - k / T &  \displaystyle  \frac{1}{ T^{3 / 2 } } \sum_{t=1}^T z_{2t}^{\prime}
\\
\\
\displaystyle \displaystyle \frac{1}{ T^{ \frac{1}{2} + \delta } } \sum_{t=1}^T z_{2t} & \displaystyle \frac{1}{ T^{1 + \delta } } \sum_{t=1}^T z_{2t} z_{2t}^{\prime}
\end{bmatrix}
\Rightarrow
\begin{bmatrix}
1 - \pi_0 & 0
\\
\\
\underline{J}_{c}(1) - \underline{J}_{c}(\pi_0)    &  (1 - \pi_0) V_{zz}
\end{bmatrix}
\end{align}

\newpage 

Furthermore, for each estimator we have that 
\begin{align}
\mathcal{D}_1  \left(\widehat{\theta}_1 - \theta \right) 
&= 
\begin{bmatrix}
k / T &  \displaystyle  \frac{1}{ T^{ 3 / 2 } } \sum_{t=1}^T  x^{\prime}_{1t}   
\\
\\
\displaystyle \displaystyle \frac{1}{ T^{ \frac{1}{2} + \delta}} \sum_{t=1}^T z_{1t}   & \displaystyle \frac{1}{ T^{1 + \delta } } \sum_{t=1}^T z_{1t} x_{1t}^{\prime}
\end{bmatrix}^{-1}
\times
\begin{bmatrix}
\displaystyle \frac{1}{ T^{ 1 - \frac{\delta}{2} } } \sum_{t=1}^T u_{t} I_{1t}
\nonumber
\\
\displaystyle \frac{1}{ T^{ \frac{1 +  \delta}{2} } } \sum_{t=1}^T z_{1t} u_t
\end{bmatrix}
\\
&\Rightarrow
\begin{bmatrix}
\pi_0 & \ \ \displaystyle  0
\\
\\
\underline{J}_c( \pi_0 )   & \ \  \displaystyle - \pi_0 \Omega_{vv} 
\end{bmatrix}^{-1}
\times
\begin{bmatrix}
0
\\
\underline{B}( \pi_0 )
\end{bmatrix}
\\
\nonumber
\\
\mathcal{D}_1 \left(\widehat{\theta}_2 - \theta \right) 
&= 
\begin{bmatrix}
1 - k / T &  \displaystyle  \frac{1}{ T^{ 3 / 2 } } \sum_{t=1}^T  x^{\prime}_{2t}   
\\
\\
\displaystyle \displaystyle \frac{1}{ T^{ \frac{1}{2} + \delta}} \sum_{t=1}^T z_{2t}   & \displaystyle \frac{1}{ T^{1 + \delta } } \sum_{t=1}^T z_{2t} x_{2t}^{\prime}
\end{bmatrix}
^{-1}
\times
\begin{bmatrix}
\displaystyle \frac{1}{ T^{ 1 - \frac{\delta}{2} } } \sum_{t=1}^T u_{t} I_{2t}
\\
\displaystyle \frac{1}{ T^{ \frac{1 +  \delta}{2} } } \sum_{t=1}^T z_{2t} u_t
\end{bmatrix}
\nonumber
\\
&\Rightarrow 
\begin{bmatrix}
(1 - \pi_0) & \ \ 0
\\
\\
\underline{J}_c( 1 ) - \underline{J}_c( \pi_0 )  & \ \  \displaystyle - (1 -\pi_0) \Omega_{vv} 
\end{bmatrix}^{-1}
\times
\begin{bmatrix}
0
\\
\underline{B}( 1 ) -  \underline{B}( \pi_0 )
\end{bmatrix}
\end{align}

\color{blue}
We have the following formula for the inverse of a partitioned matrix
\begin{align*}
\begin{pmatrix}
A_{11} & A_{12}  \\
A_{21} & A_{22} 
\end{pmatrix}^{-1}
=
\begin{pmatrix}
\left( A_{11} - A_{12} A_{22}^{-1} A_{21}  \right)^{-1}  & - A_{11}^{-1} A_{12} \mathcal{S}^{-1}   
\\
\\
- \mathcal{S}^{-1} A_{21} A_{11}^{-1}     & \mathcal{S}^{-1}
\end{pmatrix}
\end{align*}
where
\begin{align*}
\mathcal{S} = \left( A_{22} - A_{21} A_{11}^{-1} A_{12} \right)^{-1} 
\end{align*}
\color{black}
Thus, for the inversion of $\mathcal{A}_1$ we have that
\begin{align}
\mathcal{A}_1^{-1} :=  
\begin{bmatrix}
\pi_0 & \ \ 0
\\
\\
\underline{J}_c( \pi_0 )   & \ \  \displaystyle - \pi_0 \Omega_{vv} 
\end{bmatrix}^{-1}
=
\begin{bmatrix}
\displaystyle \frac{1}{\pi_0} & \ \ 0
\\
\\
\displaystyle   \frac{1}{\pi_0} \Omega_{vv}^{-1} \underline{J}_c( \pi_0 )  & \ \  \displaystyle - \frac{1}{\pi_0} \Omega_{vv}^{-1}
\end{bmatrix}
\end{align}
since
\begin{align}
\mathcal{S}^{-1} =  \left( - \pi_0 \Omega_{vv}  \right)^{-1} = - \frac{1}{\pi_0} \Omega_{vv}^{-1} 
\end{align}

\newpage 

Similarly, for the inversion of $\mathcal{A}_2$ we have that
\begin{align}
\mathcal{A}_2^{-1} &:=  
\begin{bmatrix}
1 - \pi_0 & \ \ 0
\\
\\
\underline{J}_c( 1 ) - \underline{J}_c( \pi_0 )   & \ \  \displaystyle - (1 - \pi_0 ) \Omega_{vv} 
\end{bmatrix}^{-1}
\nonumber
\\
\nonumber
\\
&=
\begin{bmatrix}
\displaystyle \frac{1}{1 - \pi_0} & \ \ 0
\\
\\
 \displaystyle  \frac{1}{(1 - \pi_0)^2} \Omega_{vv}^{-1} \bigg( \underline{J}_c( 1 ) - \underline{J}_c( \pi_0 ) \bigg) & \ \  \displaystyle - (1 - \pi_0 ) \Omega_{vv} 
\end{bmatrix}
\end{align}
Furthermore, for each estimator we have that 
\begin{align}
\mathcal{D}_1  \left(\widehat{\theta}_1 - \theta \right) 
&\Rightarrow
\begin{bmatrix}
\pi_0 & \ \ \displaystyle  0
\\
\\
\underline{J}_c( \pi_0 )   & \ \  \displaystyle - \pi_0 \Omega_{vv} 
\end{bmatrix}^{-1}
\times
\begin{bmatrix}
0
\\
\underline{B}( \pi_0 )
\end{bmatrix}
\nonumber
\\
&= 
\begin{bmatrix}
\displaystyle \frac{1}{\pi_0} & \ \ 0
\\
\\
\displaystyle \frac{1}{\pi_0^2} \Omega_{vv}^{-1} \underline{J}_c( \pi_0 )  & \ \  \displaystyle - \frac{1}{\pi_0} \Omega_{vv}^{-1}
\end{bmatrix}
\times
\begin{bmatrix}
0
\\
\underline{B}( \pi_0 )
\end{bmatrix}
\nonumber
\\
\nonumber
\\
&= 
\begin{bmatrix}
0
\\
- \displaystyle \frac{1}{\pi_0} \Omega_{vv}^{-1} \underline{B}( \pi_0 )
\end{bmatrix}
\\
\nonumber
\\
\mathcal{D}_1 \left(\widehat{\theta}_2 - \theta \right) 
&\Rightarrow 
\begin{bmatrix}
(1 - \pi_0) & \ \ 0
\\
\\
\underline{J}_c( 1 ) - \underline{J}_c( \pi_0 )  & \ \  \displaystyle - (1 -\pi_0) \Omega_{vv} 
\end{bmatrix}^{-1}
\times
\begin{bmatrix}
0
\\
\underline{B}( 1 ) -  \underline{B}( \pi_0 )
\end{bmatrix}
\nonumber
\\
&= 
\begin{bmatrix}
\displaystyle \frac{1}{1 - \pi_0} & \ \ 0
\\
\\
\displaystyle \frac{1}{(1 - \pi_0)^2} \Omega_{vv}^{-1} \bigg( \underline{J}_c( 1 ) - \underline{J}_c( \pi_0 ) \bigg) & \ \  \displaystyle - \displaystyle \frac{1}{1 - \pi_0} \Omega_{vv}^{-1} 
\end{bmatrix}
\times
\begin{bmatrix}
0
\\
\underline{B}( 1 ) -  \underline{B}( \pi_0 )
\end{bmatrix}
\nonumber
\\
\nonumber
\\
&= 
\begin{bmatrix}
0
\\
- \displaystyle \frac{1}{1 - \pi_0} \Omega_{vv}^{-1}  \bigg( \underline{B}( 1 ) -  \underline{B}( \pi_0 ) \bigg)
\end{bmatrix}
\end{align}

\newpage 

\begin{landscape}
\begin{align*}
\mathcal{D}_1 \mathcal{Q}_{ \mathcal{R} } \mathcal{D}_2   
&=  
\bigg[ \mathcal{D}_2^{-1} \left(Z_1^{\prime} X_1 \right) \mathcal{D}_1^{-1} \bigg]^{-1}  \bigg[ \mathcal{D}_2^{-1} \left( Z_1^{\prime} Z_1 \right) \mathcal{D}_1^{-1} \bigg] \bigg[ \mathcal{D}_2^{-1} \left(Z_1^{\prime} X_1 \right) \mathcal{D}_1^{-1} \bigg]^{-1}
\nonumber
\\
&\ +  
\bigg[ \mathcal{D}_2^{-1} \left(Z_2^{\prime} X_2 \right) \mathcal{D}_1^{-1} \bigg]^{-1}  \bigg[ \mathcal{D}_2^{-1} \left( Z_2^{\prime} Z_2 \right) \mathcal{D}_1^{-1} \bigg] \bigg[ \mathcal{D}_2^{-1} \left(Z_2^{\prime} X_2 \right) \mathcal{D}_1^{-1} \bigg]^{-1}  
\\
\\
&= 
\begin{bmatrix}
\displaystyle \frac{1}{\pi_0} & \ \ 0
\\
\\
\displaystyle \frac{1}{\pi_0^2} \Omega_{vv}^{-1} \underline{J}_c( \pi_0 )  & \ \  \displaystyle -  \frac{1}{\pi_0} \Omega_{vv}^{-1}
\end{bmatrix}
\times
\begin{bmatrix}
\pi_0 &  \ \ 0
\\
\\
\underline{J}_{c} ( \pi_0 )   &  \pi_0 V_{zz}  
\end{bmatrix}
\times
\begin{bmatrix}
\displaystyle \frac{1}{\pi_0} & \ \ 0
\\
\\
\displaystyle  \frac{1}{\pi_0^2} \Omega_{vv}^{-1} \underline{J}_c( \pi_0 )  & \ \  \displaystyle - \frac{1}{\pi_0} \Omega_{vv}^{-1}
\end{bmatrix}
\\
\\
&+ 
\begin{bmatrix}
\displaystyle \frac{1}{1 - \pi_0} & \ \ 0
\\
\\
\displaystyle  \frac{1}{(1 - \pi_0)^2} \Omega_{vv}^{-1} \bigg( \underline{J}_c( 1 ) - \underline{J}_c( \pi_0 ) \bigg)  & \ \  \displaystyle - \frac{1}{1 - \pi_0} \Omega_{vv}^{-1}
\end{bmatrix}
\times
\begin{bmatrix}
1 - \pi_0 &  \ \ 0
\\
\\
\underline{J}_{c} ( 1 ) - \underline{J}_{c} ( \pi_0 )   &  (1 - \pi_0) V_{zz}  
\end{bmatrix}
\times
\\
&\times 
\begin{bmatrix}
\displaystyle \frac{1}{1 - \pi_0} & \ \ 0
\\
\\
\displaystyle  \frac{1}{(1 - \pi_0)^2} \Omega_{vv}^{-1} \bigg( \underline{J}_c( 1 ) - \underline{J}_c( \pi_0 ) \bigg) & \ \  \displaystyle - \frac{1}{1 - \pi_0} \Omega_{vv}^{-1}
\end{bmatrix}
\end{align*}

\newpage 

\begin{align}
\Phi_1 
&:= 
\begin{bmatrix}
\displaystyle \frac{1}{\pi_0} & \ \ 0
\\
\\
\displaystyle  \frac{1}{\pi_0^2} \Omega_{vv}^{-1} \underline{J}_c( \pi_0 )  & \ \  \displaystyle - \frac{1}{\pi_0} \Omega_{vv}^{-1}
\end{bmatrix}
\times
\begin{bmatrix}
\pi_0 &  \ \ 0
\\
\\
\underline{J}_{c} ( \pi_0 )   &  \pi_0 V_{zz}  
\end{bmatrix}
\times
\begin{bmatrix}
\displaystyle \frac{1}{\pi_0} & \ \ 0
\\
\\
\displaystyle  \frac{1}{\pi_0^2} \Omega_{vv}^{-1} \underline{J}_c( \pi_0 )  & \ \  \displaystyle - \frac{1}{\pi_0} \Omega_{vv}^{-1} 
\end{bmatrix}
\nonumber
\\
\nonumber
\\
&=
\begin{bmatrix}
\displaystyle 1 & \ \ 0
\\
\\
\displaystyle  \frac{1}{\pi_0} \Omega_{vv}^{-1} \underline{J}_c( \pi_0 ) - \frac{1}{\pi_0} \Omega_{vv}^{-1}  \underline{J}_c( \pi_0 )  & \ \  \displaystyle - \Omega_{vv}^{-1} V_{zz}
\end{bmatrix}
\times 
\begin{bmatrix}
\displaystyle \frac{1}{\pi_0} & \ \ 0
\\
\\
\displaystyle  \frac{1}{\pi_0^2} \Omega_{vv}^{-1} \underline{J}_c( \pi_0 )  & \ \  \displaystyle -  \frac{1}{\pi_0} \Omega_{vv}^{-1} 
\end{bmatrix}
\nonumber
\\
\nonumber
\\
&=
\begin{bmatrix}
\displaystyle \frac{1}{\pi_0} & \ \ 0
\\
\\
\displaystyle \left( \Phi_1 \right)_{21}    & \ \  \displaystyle \frac{1}{\pi_0} \Omega_{vv}^{-1} V_{zz} \Omega_{vv}^{-1} 
\end{bmatrix}
\end{align}

where
\begin{align}
\left( \Phi_1 \right)_{21} =  - \frac{1}{\pi_0^2} \Omega_{vv}^{-1}V_{zz} \Omega_{vv}^{-1} \underline{J}_c( \pi_0 ) 
\end{align}

\newpage 

\begin{align*}
\Phi_2 
&= 
\begin{bmatrix}
\displaystyle \frac{1}{1 - \pi_0} & \ 0
\\
\\
\displaystyle  \frac{1}{(1 - \pi_0)^2} \Omega_{vv}^{-1} \bigg( \underline{J}_c( 1 ) - \underline{J}_c( \pi_0 ) \bigg) & \displaystyle - \frac{1}{1 - \pi_0} \Omega_{vv}^{-1}
\end{bmatrix}
\begin{bmatrix}
1 - \pi_0 &  \ 0
\\
\\
\underline{J}_{c} ( 1 ) - \underline{J}_{c} ( \pi_0 )   &  (1 - \pi_0) V_{zz}  
\end{bmatrix}
\begin{bmatrix}
\displaystyle \frac{1}{1 - \pi_0} & \ \ 0
\\
\\
\displaystyle  \frac{1}{(1 - \pi_0)^2} \Omega_{vv}^{-1} \bigg( \underline{J}_c( 1 ) - \underline{J}_c( \pi_0 ) \bigg) & \displaystyle - \frac{1}{1 - \pi_0} \Omega_{vv}^{-1}
\end{bmatrix}
\\
\nonumber
\\
&= 
\begin{bmatrix}
\displaystyle 1 & \ 0
\\
\\
\displaystyle \frac{1}{1 - \pi_0} \Omega_{vv}^{-1} \bigg( \underline{J}_c( 1 ) - \underline{J}_c( \pi_0 ) \bigg) - \frac{1}{1 - \pi_0} \Omega_{vv}^{-1} \bigg( \underline{J}_c( 1 ) - \underline{J}_c( \pi_0 ) \bigg) & \displaystyle -  \Omega_{vv}^{-1}  V_{zz}
\end{bmatrix}
\times
\begin{bmatrix}
\displaystyle \frac{1}{1 - \pi_0} & \ \ 0
\\
\\
\displaystyle \frac{1}{(1 - \pi_0)^2} \Omega_{vv}^{-1} \bigg( \underline{J}_c( 1 ) - \underline{J}_c( \pi_0 ) \bigg) & \displaystyle - \frac{1}{1 - \pi_0} \Omega_{vv}^{-1}
\end{bmatrix}
\\
\nonumber
\\
&=
\begin{bmatrix}
\displaystyle \frac{1}{1 - \pi_0}  & \ 0
\\
\\
\left( \Phi_2 \right)_{21} & \displaystyle  \frac{1}{1 - \pi_0}  \Omega_{vv}^{-1} V_{zz} \Omega_{vv}^{-1}
\end{bmatrix}
\end{align*}

where
\begin{align}
\left( \Phi_2 \right)_{21} =  - \frac{1}{(1 - \pi_0)^2} \Omega_{vv}^{-1}  V_{zz} \Omega_{vv}^{-1}  \bigg( \underline{J}_c( 1 ) - \underline{J}_c( \pi_0 ) \bigg) 
\end{align}

\newpage 

Thus, we have that
\begin{align}
\mathcal{D}_1 \mathcal{Q}_{ \mathcal{R} } \mathcal{D}_2 = \Phi_1  + \Phi_2 
&=
\begin{bmatrix}
\displaystyle \frac{1}{\pi_0} \ \ & \ \ 0
\\
\\
\left( \Phi_1 \right)_{21} \ \  & \ \  \displaystyle \frac{1}{\pi_0} \Omega_{vv}^{-1} V_{zz} \Omega_{vv}^{-1} 
\end{bmatrix}
+
\begin{bmatrix}
\displaystyle \frac{1}{1 - \pi_0} \ \ & \ \ 0
\\
\\
\displaystyle \left( \Phi_2 \right)_{21} \ \ &  \ \ \displaystyle  \frac{1}{1 - \pi_0}  \Omega_{vv}^{-1} V_{zz} \Omega_{vv}^{-1}
\end{bmatrix}
\nonumber
\\
\nonumber
\\
&=
\begin{bmatrix}
\displaystyle \frac{1}{\pi_0 (1 - \pi_0)} \ \ & \ \ 0
\\
\\
\displaystyle \left( \Phi_1 \right)_{21} +  \left( \Phi_2 \right)_{21} \ \ & \ \ \displaystyle \frac{1}{\pi_0 (1 - \pi_0)} \Omega_{vv}^{-1} V_{zz} \Omega_{vv}^{-1}
\end{bmatrix}
\end{align}
where
\begin{align}
\left( \Phi_1 \right)_{21} + \left( \Phi_2 \right)_{21} 
&= 
- \frac{1}{\pi_0^2} \Omega_{vv}^{-1}V_{zz} \Omega_{vv}^{-1} \underline{J}_c( \pi_0 )
- \frac{1}{(1 - \pi_0)^2} \Omega_{vv}^{-1}  V_{zz} \Omega_{vv}^{-1}  \bigg( \underline{J}_c( 1 ) - \underline{J}_c( \pi_0 ) \bigg) 
\nonumber
\\
\nonumber
\\
&=
- \frac{1}{\pi_0^2} \Omega_{vv}^{-1}V_{zz} \Omega_{vv}^{-1} \underline{J}_c( \pi_0 ) 
- \frac{1}{(1 - \pi_0)^2} \Omega_{vv}^{-1}  V_{zz} \Omega_{vv}^{-1} \underline{J}_c( 1 )
+ \frac{1}{(1 - \pi_0)^2} \Omega_{vv}^{-1}  V_{zz} \Omega_{vv}^{-1} \underline{J}_c( \pi_0 )
\nonumber
\\
\nonumber
\\
&=
\frac{1 - 2 \pi_0 }{\pi_0^2(1 - \pi_0)^2 } \Omega_{vv}^{-1}V_{zz} \Omega_{vv}^{-1} \underline{J}_c( \pi_0 ) 
- \frac{\pi_0^2}{\pi_0^2(1 - \pi_0)^2} \Omega_{vv}^{-1}  V_{zz} \Omega_{vv}^{-1} \underline{J}_c( 1 )
\nonumber
\\
\nonumber
\\
&=
\frac{1}{\pi_0^2(1 - \pi_0)^2 }   \Omega_{vv}^{-1}V_{zz} \Omega_{vv}^{-1} \bigg\{ \left( 1 - 2 \pi_0  \right)  \underline{J}_c( \pi_0 ) -  \pi_0^2    \underline{J}_c( 1 ) \bigg\} := \Delta
\end{align}

\newpage 

Furthermore, we have that

\begin{align}
\bigg[ \mathcal{D}_1 \mathcal{Q}_{ \mathcal{R} } \mathcal{D}_2 \bigg]^{-1}
&\equiv
\begin{bmatrix}
\displaystyle \frac{1}{\pi_0 (1 - \pi_0)} \ \ & \ \ 0
\\
\\
\displaystyle \Delta \ \ & \ \ \displaystyle \frac{1}{\pi_0 (1 - \pi_0)} \Omega_{vv}^{-1} V_{zz} \Omega_{vv}^{-1}
\end{bmatrix}^{-1}
\nonumber
\\
\nonumber
\\
&=
\begin{bmatrix}
\displaystyle \pi_0 (1 - \pi_0) \ \ & \ \ 0
\\
\\
\displaystyle K \ \ & \ \ \displaystyle \pi_0 (1 - \pi_0) \Omega_{vv} V_{zz}^{-1} \Omega_{vv}
\end{bmatrix}
\end{align}
where 
\begin{align*}
K 
&=
- \pi_0^2 (1 - \pi_0)^2 \Omega_{vv} V_{zz}^{-1} \Omega_{vv} \times \Delta
\nonumber
\\
\nonumber
\\
&= - \pi_0^2 (1 - \pi_0)^2 \Omega_{vv} V_{zz}^{-1} \Omega_{vv} \times \frac{1}{\pi_0^2(1 - \pi_0)^2 }   \Omega_{vv}^{-1}V_{zz} \Omega_{vv}^{-1} \bigg\{ \left( 1 - 2 \pi_0  \right)  \underline{J}_c( \pi_0 ) -  \pi_0^2    \underline{J}_c( 1 ) \bigg\}
\nonumber
\\
\nonumber
\\
&= 
- \bigg\{ \left( 1 - 2 \pi_0  \right)  \underline{J}_c( \pi_0 ) -  \pi_0^2    \underline{J}_c( 1 ) \bigg\}
\end{align*}

\color{black}

\end{landscape}

\newpage 

\begin{landscape} 
 
Also, the statistical distance measure is given by
\begin{align}
\mathcal{D}_1  \left( \widehat{\theta}_1 - \widehat{\theta}_2 \right) 
=
\begin{bmatrix}
0
\\
\\
-  \displaystyle  \Omega_{vv}^{-1} \left\{ \frac{1}{\pi_0} \underline{B}( \pi_0 ) - \frac{1}{1 - \pi_0} \bigg( \underline{B}( 1 ) -  \underline{B}( \pi_0 ) \bigg)  \right\}
\end{bmatrix}
=
\begin{bmatrix}
0
\\
\\
- \displaystyle \frac{1}{\pi_0( 1 - \pi_0)}  \Omega_{vv}^{-1} \bigg\{ \underline{B}( \pi_0)  - \pi_0 \underline{B}( 1 ) \bigg\}
\end{bmatrix}
\end{align}
 
Thus, the Wald IVX statistic for the case of mildly integrated regressors becomes
\begin{align}
\mathcal{W}_T( \pi)  
&\Rightarrow 
\frac{1}{\sigma_u^2}
\color{blue}
\begin{bmatrix}
a_1
\\
\\
a_2
\end{bmatrix}^{\prime} 
\color{black}
\begin{bmatrix}
\displaystyle \pi_0 (1 - \pi_0)  & \ 0
\\
\\
\textcolor{red}{K} & \displaystyle \pi_0 (1 - \pi_0) \Omega_{vv} V_{zz}^{-1} \Omega_{vv}
\end{bmatrix}
\begin{bmatrix}
0
\\
\\
- \displaystyle \frac{1}{\pi_0( 1 - \pi_0)}  \Omega_{vv}^{-1} \bigg\{ \underline{B}( \pi_0)  - \pi_0 \underline{B}( 1 ) \bigg\}
\end{bmatrix}
\end{align}

\color{red}
\underline{Note:} See next Section, for the asymptotic convergence of the blue vector above (since it has a different normalization matrix).  
\color{black}


\end{landscape} 

\newpage 

\subsubsection{Statistical Distance measure with $\mathcal{D}_2$ normalization matrix}

We have that
\begin{align*}
\mathcal{D}_2 \left( \hat{\theta}_1 - \theta  \right) 
&= \left[ \mathcal{D}_1^{-1} \left( {Z}_1^{\prime} X_1 \right) \mathcal{D}_2^{-1} \right]^{-1} \mathcal{D}_1^{-1} \left( {Z}_1^{\prime} u \right)
\\
\mathcal{D}_2 \left( \hat{\theta}_2 - \theta  \right) 
&= \left[ \mathcal{D}_1^{-1} \left( {Z}_2^{\prime} X_2 \right) \mathcal{D}_2^{-1} \right]^{-1} \mathcal{D}_1^{-1} \left( {Z}_2^{\prime} u \right)
\end{align*}
Therefore, 
\begin{align*}
\mathcal{D}_2 \left( \hat{\theta}_1 - \hat{\theta}_2  \right) = \left\{ \left[ \mathcal{D}_1^{-1} \left( {Z}_1^{\prime} X_1 \right) \mathcal{D}_2^{-1} \right]^{-1} \mathcal{D}_1^{-1} \left( {Z}_1^{\prime} u \right) - \left[ \mathcal{D}_1^{-1} \left( {Z}_2^{\prime} X_2 \right) \mathcal{D}_2^{-1} \right]^{-1} \mathcal{D}_1^{-1} \left( {Z}_2^{\prime} u \right) \right\}
\end{align*}
Then, we consider the weakly convergence of the following sample moments
\begin{align}
\mathcal{D}_1^{-1} \left( Z_1^{\prime} X_1 \right) \mathcal{D}_2^{-1} 
&= 
\begin{bmatrix}
k / T &  \displaystyle  \frac{1}{ T^{ \frac{1}{2} + \delta}} \sum_{t=1}^T  x^{\prime}_{1t}   
\\
\nonumber
\\
\displaystyle \displaystyle \frac{1}{ T^{ 3 / 2 } }  \sum_{t=1}^T z_{1t}   & \displaystyle \frac{1}{ T^{1 + \delta } } \sum_{t=1}^T z_{1t} x_{1t}^{\prime}
\end{bmatrix}
\Rightarrow
\begin{bmatrix}
\pi_0 & \ \ \displaystyle  \underline{J}_c( \pi_0 )
\\
\\
0 & \ \ \displaystyle - \pi_0 \Omega_{vv}
\end{bmatrix} 
\\
\\
\mathcal{D}_1^{-1} \left( Z_2^{\prime} X_2 \right) \mathcal{D}_2^{-1} 
&= 
\begin{bmatrix}
1 - k / T &  \displaystyle  \frac{1}{ T^{ \frac{1}{2} + \delta}}  \sum_{t=1}^T  x^{\prime}_{2t}   
\\
\\
\displaystyle \frac{1}{ T^{ 3 / 2 } } \sum_{t=1}^T z_{2t}   & \displaystyle \frac{1}{ T^{1 + \delta } } \sum_{t=1}^T z_{2t} x_{2t}^{\prime}
\end{bmatrix}
\Rightarrow
\begin{bmatrix}
(1 - \pi_0) & \ \ \displaystyle \underline{J}_c( 1 ) - \underline{J}_c( \pi_0 ) 
\\
\\
0 & \ \  \displaystyle - (1 - \pi_0) \Omega_{vv}
\end{bmatrix} 
\end{align}
\begin{align}
\mathcal{D}_1^{-1} \left( \sum_{t=1}^T Z_{1}^{\prime} Z_1 \right) \mathcal{D}_2^{-1} = \begin{bmatrix}
k / T &  \displaystyle  \frac{1}{ T^{ \frac{1}{2} + \delta}}  \sum_{t=1}^T z_{1t}^{\prime}
\\
\\
\displaystyle \frac{1}{ T^{3 / 2 } }\sum_{t=1}^T z_{1t} & \displaystyle \frac{1}{ T^{1 + \delta } } \sum_{t=1}^T z_{1t} z_{1t}^{\prime}
\end{bmatrix}
\Rightarrow
\begin{bmatrix}
\pi_0 &  \ \ \underline{J}_{c} ( \pi_0 ) 
\\
\\
0  &  \pi_0 V_{zz}  
\end{bmatrix}
\end{align}
\begin{align}
\mathcal{D}_1^{-1} \left( \sum_{t=1}^T Z_{2}^{\prime} Z_2 \right) \mathcal{D}_2^{-1} 
= \begin{bmatrix}
1 - k / T &  \displaystyle  \frac{1}{ T^{ \frac{1}{2} + \delta } }  \sum_{t=1}^T z_{2t}^{\prime}
\\
\\
\displaystyle \frac{1}{ T^{3 / 2 } } \sum_{t=1}^T z_{2t} & \displaystyle \frac{1}{ T^{1 + \delta } } \sum_{t=1}^T z_{2t} z_{2t}^{\prime}
\end{bmatrix}
\Rightarrow
\begin{bmatrix}
1 - \pi_0 & \underline{J}_{c}(1) - \underline{J}_{c}(\pi_0) 
\\
\\
0   &  (1 - \pi_0) V_{zz}
\end{bmatrix}
\end{align}

\newpage 

Furthermore, for each estimator we have that 
\begin{align}
\mathcal{D}_2  \left(\widehat{\theta}_1 - \theta \right) 
&= 
\begin{bmatrix}
k / T &  \displaystyle \frac{1}{ T^{ \frac{1}{2} + \delta}}   \sum_{t=1}^T  x^{\prime}_{1t}   
\\
\\
\displaystyle \frac{1}{ T^{ 3 / 2 } }  \sum_{t=1}^T z_{1t}   & \displaystyle \frac{1}{ T^{1 + \delta } } \sum_{t=1}^T z_{1t} x_{1t}^{\prime}
\end{bmatrix}^{-1}
\times
\begin{bmatrix}
\displaystyle \frac{1}{ T^{ \frac{\delta}{2} } } \sum_{t=1}^T u_{t} I_{1t}
\nonumber
\\
\displaystyle \frac{1}{ T^{ \frac{1 +  \delta}{2} } } \sum_{t=1}^T z_{1t} u_t
\end{bmatrix}
\\
&\Rightarrow
\begin{bmatrix}
\pi_0 & \ \ \displaystyle  \underline{J}_c( \pi_0 )
\\
\\
0   & \ \  \displaystyle - \pi_0 \Omega_{vv} 
\end{bmatrix}^{-1}
\times
\begin{bmatrix}
0
\\
\underline{B}( \pi_0 )
\end{bmatrix}
\\
\nonumber
\\
\mathcal{D}_2 \left(\widehat{\theta}_2 - \theta \right) 
&= 
\begin{bmatrix}
1 - k / T &  \displaystyle  \frac{1}{ T^{ \frac{1}{2} + \delta}}  \sum_{t=1}^T  x^{\prime}_{2t}   
\\
\\
\displaystyle \frac{1}{ T^{ 3 / 2 } }  \sum_{t=1}^T z_{2t}   & \displaystyle \frac{1}{ T^{1 + \delta } } \sum_{t=1}^T z_{2t} x_{2t}^{\prime}
\end{bmatrix}
^{-1}
\times
\begin{bmatrix}
\displaystyle \frac{1}{ T^{ \frac{\delta}{2} } } \sum_{t=1}^T u_{t} I_{2t}
\\
\displaystyle \frac{1}{ T^{ \frac{1 +  \delta}{2} } } \sum_{t=1}^T z_{2t} u_t
\end{bmatrix}
\nonumber
\\
&\Rightarrow 
\begin{bmatrix}
(1 - \pi_0) & \ \ \underline{J}_c( 1 ) - \underline{J}_c( \pi_0 )
\\
\\
 0 & \ \  \displaystyle - (1 -\pi_0) \Omega_{vv} 
\end{bmatrix}^{-1}
\times
\begin{bmatrix}
0
\\
\underline{B}( 1 ) -  \underline{B}( \pi_0 )
\end{bmatrix}
\end{align}

\color{blue}
We have the following formula for the inverse of a partitioned matrix
\begin{align*}
\begin{pmatrix}
A_{11} & A_{12}  \\
A_{21} & A_{22} 
\end{pmatrix}^{-1}
=
\begin{pmatrix}
\left( A_{11} - A_{12} A_{22}^{-1} A_{21}  \right)^{-1}  & - A_{11}^{-1} A_{12} \mathcal{S}^{-1}   
\\
\\
- \mathcal{S}^{-1} A_{21} A_{11}^{-1}     & \mathcal{S}^{-1}
\end{pmatrix}
\end{align*}
where
\begin{align*}
\mathcal{S} = \left( A_{22} - A_{21} A_{11}^{-1} A_{12} \right)^{-1} 
\end{align*}
\color{black}
Thus, for the inversion of $\mathcal{A}_1$ we have that
\begin{align}
\mathcal{A}_1^{-1} :=  
\begin{bmatrix}
\pi_0 & \ \ \underline{J}_c( \pi_0 )
\\
\\
0   & \ \  \displaystyle - \pi_0 \Omega_{vv} 
\end{bmatrix}^{-1}
=
\begin{bmatrix}
\displaystyle \frac{1}{\pi_0} & \ \ \displaystyle  \frac{1}{\pi_0} \Omega_{vv}^{-1} \underline{J}_c( \pi_0 )
\\
\\
0    & \ \  \displaystyle - \frac{1}{\pi_0} \Omega_{vv}^{-1}
\end{bmatrix}
\end{align}
since
\begin{align}
\mathcal{S}^{-1} =  \left( - \pi_0 \Omega_{vv}  \right)^{-1} = - \frac{1}{\pi_0} \Omega_{vv}^{-1} 
\end{align}

\newpage 

Similarly, for the inversion of $\mathcal{A}_2$ we have that
\begin{align}
\mathcal{A}_2^{-1} &:=  
\begin{bmatrix}
1 - \pi_0 & \ \ \underline{J}_c( 1 ) - \underline{J}_c( \pi_0 )
\\
\\
0   & \ \  \displaystyle - (1 - \pi_0 ) \Omega_{vv} 
\end{bmatrix}^{-1}
\nonumber
\\
\nonumber
\\
&=
\begin{bmatrix}
\displaystyle \frac{1}{1 - \pi_0} & \ \  \displaystyle  \frac{1}{(1 - \pi_0)^2} \Omega_{vv}^{-1} \bigg( \underline{J}_c( 1 ) - \underline{J}_c( \pi_0 ) \bigg)
\\
\\
0 & \ \  \displaystyle - (1 - \pi_0 ) \Omega_{vv} 
\end{bmatrix}
\end{align}
Furthermore, for each estimator we have that 
\begin{align}
\mathcal{D}_2  \left(\widehat{\theta}_1 - \theta \right) 
&\Rightarrow
\begin{bmatrix}
\pi_0 & \ \ \displaystyle  \underline{J}_c( \pi_0 ) 
\\
\\
0 & \ \  \displaystyle - \pi_0 \Omega_{vv} 
\end{bmatrix}^{-1}
\times
\begin{bmatrix}
0
\\
\underline{B}( \pi_0 )
\end{bmatrix}
\nonumber
\\
&= 
\begin{bmatrix}
\displaystyle \frac{1}{\pi_0} & \ \ \displaystyle \frac{1}{\pi_0^2} \Omega_{vv}^{-1} \underline{J}_c( \pi_0 )
\\
\\
0 & \ \  \displaystyle - \frac{1}{\pi_0} \Omega_{vv}^{-1}
\end{bmatrix}
\times
\begin{bmatrix}
0
\\
\underline{B}( \pi_0 )
\end{bmatrix}
\nonumber
\\
\nonumber
\\
&= 
\begin{bmatrix}
\displaystyle \frac{1}{\pi_0^2} \Omega_{vv}^{-1} \underline{J}_c( \pi_0 ) \underline{B}( \pi_0 )
\\
\\
\displaystyle - \frac{1}{\pi_0} \Omega_{vv}^{-1}  \underline{B}( \pi_0 )
\end{bmatrix}
\\
\nonumber
\\
\mathcal{D}_2 \left(\widehat{\theta}_2 - \theta \right) 
&\Rightarrow 
\begin{bmatrix}
(1 - \pi_0) & \ \ 0
\\
\\
\underline{J}_c( 1 ) - \underline{J}_c( \pi_0 )  & \ \  \displaystyle - (1 -\pi_0) \Omega_{vv} 
\end{bmatrix}^{-1}
\times
\begin{bmatrix}
0
\\
\underline{B}( 1 ) -  \underline{B}( \pi_0 )
\end{bmatrix}
\nonumber
\\
&= 
\begin{bmatrix}
\displaystyle \frac{1}{1 - \pi_0} & \ \ \displaystyle \frac{1}{(1 - \pi_0)^2} \Omega_{vv}^{-1} \bigg( \underline{J}_c( 1 ) - \underline{J}_c( \pi_0 ) \bigg)
\\
\\
0 & \ \  \displaystyle - \displaystyle \frac{1}{1 - \pi_0} \Omega_{vv}^{-1} 
\end{bmatrix}
\times
\begin{bmatrix}
0
\\
\underline{B}( 1 ) -  \underline{B}( \pi_0 )
\end{bmatrix}
\nonumber
\\
\nonumber
\\
&= 
\begin{bmatrix}
\displaystyle \frac{1}{(1 - \pi_0)^2} \Omega_{vv}^{-1} \bigg( \underline{J}_c( 1 ) - \underline{J}_c( \pi_0 ) \bigg)\bigg( \underline{B}( 1 ) -  \underline{B}( \pi_0 ) \bigg)
\\
\\
- \displaystyle \frac{1}{1 - \pi_0} \Omega_{vv}^{-1}  \bigg( \underline{B}( 1 ) -  \underline{B}( \pi_0 ) \bigg)
\end{bmatrix}
\end{align}

\newpage 

\begin{landscape} 
Thus, the statistical distance measure is given by
\begin{align*}
\color{blue}
\mathcal{D}_2 \left( \widehat{\theta}_1 - \widehat{\theta}_2 \right) 
\color{black}
&=
\begin{bmatrix}
\displaystyle \frac{1}{\pi_0^2} \Omega_{vv}^{-1} \underline{J}_c( \pi_0 ) \underline{B}( \pi_0 )
\\
\\
\displaystyle - \frac{1}{\pi_0} \Omega_{vv}^{-1}  \underline{B}( \pi_0 )
\end{bmatrix}
-
\begin{bmatrix}
\displaystyle \frac{1}{(1 - \pi_0)^2} \Omega_{vv}^{-1} \bigg( \underline{J}_c( 1 ) - \underline{J}_c( \pi_0 ) \bigg)\bigg( \underline{B}( 1 ) -  \underline{B}( \pi_0 ) \bigg)
\\
\\
- \displaystyle \frac{1}{1 - \pi_0} \Omega_{vv}^{-1}  \bigg( \underline{B}( 1 ) -  \underline{B}( \pi_0 ) \bigg)
\end{bmatrix}
\\
\\
&= 
\begin{bmatrix}
\displaystyle \frac{1}{\pi_0^2} \Omega_{vv}^{-1} \underline{J}_c( \pi_0 ) \underline{B}( \pi_0 ) - \frac{1}{(1 - \pi_0)^2} \Omega_{vv}^{-1} \bigg( \underline{J}_c( 1 ) - \underline{J}_c( \pi_0 ) \bigg)\bigg( \underline{B}( 1 ) -  \underline{B}( \pi_0 ) \bigg)
\\
\\
\displaystyle - \frac{1}{\pi_0} \Omega_{vv}^{-1}  \underline{B}( \pi_0 ) +  \displaystyle \frac{1}{1 - \pi_0} \Omega_{vv}^{-1}  \bigg( \underline{B}( 1 ) -  \underline{B}( \pi_0 ) \bigg)
\end{bmatrix}
\end{align*}
Therefore, the Wald IVX statistic for the case of mildly integrated regressors becomes
\begin{align}
\mathcal{W}_T( \pi)  
&\Rightarrow 
\underset{ \pi \in [ \pi_1, \pi_2 ] }{ { \text{sup}  } } \
\frac{1}{\sigma_u^2}
\color{blue}
\bigg[ \mathcal{D}_2 \left( \widehat{\theta}_1 - \widehat{\theta}_2 \right) \bigg]^{\prime} 
\color{black}
\begin{bmatrix}
\displaystyle \pi_0 (1 - \pi_0)  & \ 0
\\
\\
\textcolor{red}{K} & \displaystyle \pi_0 (1 - \pi_0) \Omega_{vv} V_{zz}^{-1} \Omega_{vv}
\end{bmatrix}
\begin{bmatrix}
0
\\
\\
- \displaystyle \frac{1}{\pi_0( 1 - \pi_0)}  \Omega_{vv}^{-1} \bigg\{ \underline{B}( \pi_0)  - \pi_0 \underline{B}( 1 ) \bigg\}
\end{bmatrix}
\nonumber
\\
&= 
\underset{ \pi \in [ \pi_1, \pi_2 ] }{ { \text{sup}  } } \  
\frac{1}{\sigma_u^2} 
\begin{bmatrix}
\textcolor{blue}{a_1}
\\
\\
\displaystyle - V_{zz}^{-1} \Omega_{vv} \bigg( \underline{B}( \pi_0 ) - \pi_0 \underline{B}( 1 ) \bigg) 
\end{bmatrix}^{\prime}
\begin{bmatrix}
0
\\
\\
- \displaystyle \frac{1}{\pi_0( 1 - \pi_0)}  \Omega_{vv}^{-1} \bigg\{ \underline{B}( \pi_0)  - \pi_0 \underline{B}( 1 ) \bigg\}
\end{bmatrix}
\nonumber
\\
\nonumber
\\
&= 
\underset{ \pi \in [ \pi_1, \pi_2 ] }{ { \text{sup}  } } \ \frac{ \bigg[ \underline{W}( \pi) - \pi \underline{W}( 1 ) \bigg]^{\prime} \bigg[ \underline{W}( \pi) - \pi \underline{W}( 1 ) \bigg]  }{\pi (1 - \pi)}
\end{align}

\newpage 

where the elements for the first vector are as below:
\begin{align}
\textcolor{blue}{a_1}
&=
\frac{1 - \pi_0}{\pi_0} \Omega_{vv}^{-1} \underline{J}_c( \pi_0 ) \underline{B}( \pi_0 ) - \frac{1}{(1 - \pi_0)} \Omega_{vv}^{-1} \bigg( \underline{J}_c( 1 ) - \underline{J}_c( \pi_0 ) \bigg)\bigg( \underline{B}( 1 ) -  \underline{B}( \pi_0 ) \bigg)
\nonumber
\\
\nonumber
\\
&+  \left\{ \frac{1}{\pi_0} \Omega_{vv}^{-1}  \underline{B}( \pi_0 ) +  \displaystyle \frac{1}{1 - \pi_0} \Omega_{vv}^{-1}  \bigg( \underline{B}( 1 ) -  \underline{B}( \pi_0 ) \bigg) \right\}
\bigg\{ \left( 1 - 2 \pi_0  \right)  \underline{J}_c( \pi_0 ) -  \pi_0^2    \underline{J}_c( 1 ) \bigg\}
\end{align}

\begin{align}
a_2
&=  
\left\{ - \frac{1}{\pi_0} \Omega_{vv}^{-1}  \underline{B}( \pi_0 ) +  \displaystyle \frac{1}{1 - \pi_0} \Omega_{vv}^{-1}  \bigg( \underline{B}( 1 ) -  \underline{B}( \pi_0 ) \bigg)  \right\} \bigg\{ \pi_0 (1 - \pi_0) \Omega_{vv} V_{zz}^{-1} \Omega_{vv} \bigg\}
\nonumber
\\
\nonumber
\\
&- (1 - \pi_0) \Omega_{vv}^{-1} \Omega_{vv} V_{zz}^{-1} \Omega_{vv} \underline{B}( \pi_0 ) + \pi_0 \Omega_{vv}^{-1}  \Omega_{vv} V_{zz}^{-1} \Omega_{vv} \underline{B}( 1 ) - \pi_0 \Omega_{vv}^{-1}  \Omega_{vv} V_{zz}^{-1} \Omega_{vv} \underline{B}( \pi_0 )
\nonumber
\\
\nonumber
\\
&= - V_{zz}^{-1} \Omega_{vv} \bigg( \underline{B}( \pi_0 ) - \pi_0 \underline{B}( 1 ) \bigg) 
\end{align}

\end{landscape} 

\newpage 

\subsection{Wald IVX test for LUR regressors}

We have the following equivalent expression for the Wald IVX statistic
\begin{align*}
\mathcal{W}_T( \pi) 
= 
\frac{1}{ \hat{\sigma}^2 } \left( \hat{\theta}_1 - \hat{\theta}_2 \right)^{\prime} \mathcal{D}_2 \bigg[ \mathcal{D}_1  \mathcal{Q}_{ \mathcal{R} }   \mathcal{D}_2 \bigg]^{-1} \mathcal{D}_1 \left( \hat{\theta}_1 - \hat{\theta}_2 \right)
\end{align*}
To determine the limiting distribution of the Wald-IVX statistic we consider the weakly convergence of each of its terms separately. To simplify the algebra we denote with $\displaystyle  Q_{xz} := - \left( \int_{ 0 }^1 \underline{J}_c (r) d \underline{J}_v + \Omega_{vv}  \right) C_z^{-1}$. Regardless whether the break-point is known or unknown the following holds
\begin{align}
\left( \frac{1}{ T^{1+\delta}} \sum_{t=1}^k  x_{t}z_{t}^{\prime} \right)
\equiv \left( \frac{1}{ T^{1+\delta}} \sum_{t=1}^T  x_{1t}z_{1t}^{\prime}  \right) \Rightarrow \left( \int_{ 0 }^{ \pi } \underline{J}_c (r) d \underline{J}_v + \pi \Omega_{vv}  \right) C_z^{-1}
\end{align}
and
\begin{align}
\left( \frac{1}{ T^{1+\delta}} \sum_{t=k+1}^T  x_{t}z_{t}^{\prime} \right)
\equiv \left( \frac{1}{ T^{1+\delta}} \sum_{t=1}^T  x_{2t}z_{2t}^{\prime} \right) \Rightarrow \left( \int_{ \pi }^1 \underline{J}_c (r) d \underline{J}_v + (1 - \pi ) \Omega_{vv}  \right) C_z^{-1}
\end{align}

We consider the weakly convergence of the following sample moments
\begin{align}
\mathcal{D}_2^{-1} \left( Z_1^{\prime} X_1 \right) \mathcal{D}_1^{-1} 
&= 
\begin{bmatrix}
k / T &  \displaystyle  \frac{1}{ T^{ 3 / 2 } } \sum_{t=1}^T  x^{\prime}_{1t}   
\\
\nonumber
\\
\displaystyle \displaystyle \frac{1}{ T^{ \frac{1}{2} + \delta}} \sum_{t=1}^T z_{1t}   & \displaystyle \frac{1}{ T^{1 + \delta } } \sum_{t=1}^T z_{1t} x_{1t}^{\prime}
\end{bmatrix}
\Rightarrow
\begin{bmatrix}
\pi_0 & \ \ \displaystyle  \int_0^{\pi_0} \underline{J}^{\prime} (r) dr 
\\
\\
\underline{J}_c( \pi_0 ) & \ \ \displaystyle \pi_0 \Omega_{vv} + \int_{ 0 }^{\pi_0} \underline{J}_c (r) d \underline{J}_v
\end{bmatrix} 
\\
\\
\mathcal{D}_2^{-1} \left( Z_2^{\prime} X_2 \right) \mathcal{D}_1^{-1} 
&= 
\begin{bmatrix}
1 - k / T &  \displaystyle  \frac{1}{ T^{ 3 / 2 } } \sum_{t=1}^T  x^{\prime}_{2t}   
\\
\\
\displaystyle \displaystyle \frac{1}{ T^{ \frac{1}{2} + \delta}} \sum_{t=1}^T z_{2t}   & \displaystyle \frac{1}{ T^{1 + \delta } } \sum_{t=1}^T z_{2t} x_{2t}^{\prime}
\end{bmatrix}
\Rightarrow
\begin{bmatrix}
(1 - \pi_0) & \ \ \displaystyle \int_{\pi_0}^1 \underline{J}^{\prime} (r) dr 
\\
\\
\underline{J}_c( 1 ) - \underline{J}_c( \pi_0 )  & \ \  \displaystyle (1 - \pi_0) \Omega_{vv} + \int_{ \pi_0 }^{1} \underline{J}_c (r) d \underline{J}_v
\end{bmatrix} 
\end{align}
\begin{align}
\mathcal{D}_2^{-1} \left( \sum_{t=1}^T Z_{1}^{\prime} Z_1 \right) \mathcal{D}_1^{-1} = \begin{bmatrix}
k / T &  \displaystyle \frac{1}{ T^{3 / 2 } } \sum_{t=1}^T z_{1t}^{\prime}
\\
\\
\displaystyle \displaystyle \frac{1}{ T^{ \frac{1}{2} + \delta}} \sum_{t=1}^T z_{1t} & \displaystyle \frac{1}{ T^{1 + \delta } } \sum_{t=1}^T z_{1t} z_{1t}^{\prime}
\end{bmatrix}
\Rightarrow
\begin{bmatrix}
\pi_0 &  \ \ 0
\\
\\
\underline{J}_{c} ( \pi_0 )   &  \pi_0 V_{zz}  
\end{bmatrix}
\end{align}

\newpage 

\begin{align}
\mathcal{D}_2^{-1} \left( \sum_{t=1}^T Z_{2}^{\prime} Z_2 \right) \mathcal{D}_1^{-1} 
= \begin{bmatrix}
1 - k / T &  \displaystyle  \frac{1}{ T^{3 / 2 } } \sum_{t=1}^T z_{2t}^{\prime}
\\
\\
\displaystyle \displaystyle \frac{1}{ T^{ \frac{1}{2} + \delta } } \sum_{t=1}^T z_{2t} & \displaystyle \frac{1}{ T^{1 + \delta } } \sum_{t=1}^T z_{2t} z_{2t}^{\prime}
\end{bmatrix}
\Rightarrow
\begin{bmatrix}
1 - \pi_0 & 0
\\
\\
\underline{J}_{c}(1) - \underline{J}_{c}(\pi_0)    &  (1 - \pi_0) V_{zz}
\end{bmatrix}
\end{align}
Furthermore, for each estimator we have that 
\begin{align}
\mathcal{D}_1  \left(\widehat{\theta}_1 - \theta \right) 
&= 
\begin{bmatrix}
k / T &  \displaystyle  \frac{1}{ T^{ 3 / 2 } } \sum_{t=1}^T  x^{\prime}_{1t}   
\\
\\
\displaystyle \displaystyle \frac{1}{ T^{ \frac{1}{2} + \delta}} \sum_{t=1}^T z_{1t}   & \displaystyle \frac{1}{ T^{1 + \delta } } \sum_{t=1}^T z_{1t} x_{1t}^{\prime}
\end{bmatrix}^{-1}
\times
\begin{bmatrix}
\displaystyle \frac{1}{ T^{ 1 - \frac{\delta}{2} } } \sum_{t=1}^T u_{t} I_{1t}
\nonumber
\\
\displaystyle \frac{1}{ T^{ \frac{1 +  \delta}{2} } } \sum_{t=1}^T z_{1t} u_t
\end{bmatrix}
\\
&\Rightarrow
\begin{bmatrix}
\pi_0 & \ \ \displaystyle  \int_0^{\pi_0} \underline{J}^{\prime} (r) dr
\\
\\
\underline{J}_c( \pi_0 )   & \ \  \displaystyle \pi_0 \Omega_{vv} + \int_{ 0 }^{\pi_0} \underline{J}_c (r) d \underline{J}_v 
\end{bmatrix}^{-1}
\times
\begin{bmatrix}
0
\\
\underline{B}( \pi_0 )
\end{bmatrix}
\\
\nonumber
\\
\mathcal{D}_1 \left(\widehat{\theta}_2 - \theta \right) 
&= 
\begin{bmatrix}
1 - k / T &  \displaystyle  \frac{1}{ T^{ 3 / 2 } } \sum_{t=1}^T  x^{\prime}_{2t}   
\\
\\
\displaystyle \displaystyle \frac{1}{ T^{ \frac{1}{2} + \delta}} \sum_{t=1}^T z_{2t}   & \displaystyle \frac{1}{ T^{1 + \delta } } \sum_{t=1}^T z_{2t} x_{2t}^{\prime}
\end{bmatrix}
^{-1}
\times
\begin{bmatrix}
\displaystyle \frac{1}{ T^{ 1 - \frac{\delta}{2} } } \sum_{t=1}^T u_{t} I_{2t}
\\
\displaystyle \frac{1}{ T^{ \frac{1 +  \delta}{2} } } \sum_{t=1}^T z_{2t} u_t
\end{bmatrix}
\nonumber
\\
&\Rightarrow 
\begin{bmatrix}
(1 - \pi_0) & \ \ \displaystyle \int_{\pi_0}^1 \underline{J}^{\prime} (r) dr
\\
\\
\underline{J}_c( 1 ) - \underline{J}_c( \pi_0 )  & \ \  \displaystyle (1 -\pi_0) \Omega_{vv} + \int_{ \pi_0 }^{1} \underline{J}_c (r) d \underline{J}_v
\end{bmatrix}^{-1}
\times
\begin{bmatrix}
0
\\
\underline{B}( 1 ) -  \underline{B}( \pi_0 )
\end{bmatrix}
\end{align}

Note that, for the asymptotic converges of the terms $\frac{1}{ T^{ \frac{1}{2} + \delta}} \sum_{t=1}^T z_{1t}$ and $\frac{1}{ T^{ \frac{1}{2} + \delta}} \sum_{t=1}^T z_{2t}$, we use the result given by Lemma B1 (i) in the Appendix of KMS. That, is since we have that $\gamma = 1$ in the case of persistent regressors and we assume that the exponent rate of the degree of persistence of the IVX instrument is $\delta \in (0,1)$, then

\newpage 

\begin{align}
\frac{1}{ T^{ \frac{1}{2} + \delta}} \sum_{t=1}^T \tilde{z}_{t} = - C_z^{-1} \frac{1}{ T^{ \frac{1}{2} + \delta}} x_T + O_p( . ) 
\end{align}
Thus, we have that $\frac{1}{ T^{ \frac{1}{2} + \delta}} \sum_{t=1}^{ \floor{T\pi }} \tilde{z}_{t} \Rightarrow \underline{J}_c( \pi_0 )$.

Therefore, we consider the inverse of the partition of the following matrices as below
\begin{align*}
\mathcal{A}_1 :=  \bigg[ \mathcal{D}_2^{-1} \left(Z_1^{\prime} X_1 \right) \mathcal{D}_1^{-1} \bigg]^{-1}  
&= 
\begin{bmatrix}
\pi_0 & \ \ \displaystyle \int_0^{\pi_0} \underline{J}^{\prime} (r) dr
\\
\\
\underline{J}_c( \pi_0 )   & \ \  \big[ Q_{xz} \big]_{ij} ( \pi_0 ) 
\end{bmatrix}^{-1}
\nonumber
\\
\\
\mathcal{A}_2 :=  \bigg[ \mathcal{D}_2^{-1} \left(Z_2^{\prime} X_2 \right) \mathcal{D}_1^{-1} \bigg]^{-1} 
&= 
\begin{bmatrix}
(1 - \pi_0) & \ \ \displaystyle \int_{\pi_0}^1 \underline{J}^{\prime} (r) dr
\\
\\
\underline{J}_c( 1 ) - \underline{J}_c( \pi_0 )  & \ \  \big[ Q_{xz} \big]_{ij} ( 1 ) 
\end{bmatrix}^{-1}
\end{align*}

\medskip

\underline{Useful Notation:}
We use the following matrix notation to simplify further the expression for the sup Wald-IVX statistic
\begin{align}
\big[ Q_{xz} \big]_{ij} (s) := \displaystyle - \left( d(s) \Omega_{vv,ij} +  \int_{g(s)}^{s} \underline{J}_{ic} (r) d \underline{J}_{j} \right) C_z^{-1}
\end{align}
with
\begin{equation}
d(s) =
    \begin{cases}
      \pi_0,       & \text{when}\ s := \pi_0 \\
      (1 - \pi_0),   & \text{when}\ s := 1
    \end{cases}
\ \ \text{and} \ \ \ \
g(s) =
    \begin{cases}
      0,     & \text{when}\ s := \pi_0 \\
      \pi_0, & \text{when}\ s := 1
    \end{cases}
\end{equation}

\color{blue}
We have the following formula for the inverse of a partitioned matrix
\begin{align*}
\begin{pmatrix}
A_{11} & A_{12}  \\
A_{21} & A_{22} 
\end{pmatrix}^{-1}
=
\begin{pmatrix}
\left( A_{11} - A_{12} A_{22}^{-1} A_{21}  \right)^{-1}  & - A_{11}^{-1} A_{12} \mathcal{S}^{-1}   
\\
\\
- \mathcal{S}^{-1} A_{21} A_{11}^{-1}     & \mathcal{S}^{-1}
\end{pmatrix}
, \ \mathcal{S} = \left( A_{22} - A_{21} A_{11}^{-1} A_{12} \right)^{-1}
\end{align*}
\color{black}
Equivalently, we denote with
\begin{align}
\big[ Q_{xz} \big]_{ij} ( \pi_0 ) \equiv \Delta_1 
&:=  \displaystyle \pi_0 \Omega_{vv} + \int_{ 0 }^{ \pi_0 } \underline{J}_c (r) d \underline{J}_v 
\\
\big[ Q_{xz} \big]_{ij} ( 1 ) \equiv \Delta_2 
&:=  \displaystyle (1 -\pi_0) \Omega_{vv} + \int_{ \pi_0 }^{1} \underline{J}_c (r) d \underline{J}_v 
\end{align}

\newpage 

Moreover, we denote with 
\begin{align}
\mathcal{S}_1 
&= 
\left( \Delta_1  - \frac{ \underline{J}_c( \pi_0 )}{  \pi_0 } \int_0^{\pi_0} \underline{J}^{\top} (r) dr    \right)
\\
\mathcal{S}_2 &= \left( \Delta_2   - \frac{ \underline{J}_c( 1 ) - \underline{J}_c( \pi_0 )}{ 1 - \pi_0 } \int_{\pi_0}^{1} \underline{J}^{\top} (r) dr    \right)
\end{align}
Thus, for the inversion of $\mathcal{A}_1$ we have that
\begin{align}
\mathcal{A}_1^{-1} :=  
\begin{bmatrix}
\pi_0 & \ \ \displaystyle  \int_0^{\pi_0} \underline{J}^{\top} (r) dr
\\
\\
\underline{J}_c( \pi_0 )   & \ \ \Delta_1
\end{bmatrix}^{-1}
\end{align}
\begin{align*}
B_{11} &=  \left( \pi_0  - \left( \int_0^{\pi_0} \underline{J}^{\top} (r) dr \right) \Delta_1^{-1} \underline{J}_c( \pi_0 ) \right)^{-1}
\\
B_{12} &=  - \frac{1}{\pi_0} \left( \int_0^{\pi_0} \underline{J}^{\top} (r) dr \right) \mathcal{S}_1^{-1}
\\
B_{21} &=  - \mathcal{S}_1^{-1} \frac{ \underline{J}_c( \pi_0 )}{  \pi_0 }
\\
B_{22} &=  \mathcal{S}_1^{-1} 
\end{align*}
\begin{align}
\mathcal{A}_1^{-1}   \equiv 
\begin{bmatrix}
\displaystyle \left( \pi_0  - \left( \int_0^{\pi_0} \underline{J}^{\top} (r) dr \right) \Delta_1^{-1} \underline{J}_c( \pi_0 ) \right)^{-1} \ \  & \ \ \displaystyle - \frac{1}{\pi_0} \left( \int_0^{\pi_0} \underline{J}^{\top} (r) dr \right) \mathcal{S}_1^{-1}
\\
\\
\displaystyle - \mathcal{S}_1^{-1} \frac{ \underline{J}_c( \pi_0 )}{  \pi_0 }  \ \  & \ \  \displaystyle \mathcal{S}_1^{-1} 
\end{bmatrix}
\end{align}
Similarly, for $\mathcal{A}_2^{-1}$ we have that
\begin{align*}
\mathcal{A}_2 :=  \bigg[ \mathcal{D}_2^{-1} \left(Z_2^{\prime} X_2 \right) \mathcal{D}_1^{-1} \bigg]^{-1} 
&= 
\begin{bmatrix}
(1 - \pi_0) & \ \ \displaystyle \int_{\pi_0}^1 \underline{J}^{\prime} (r) dr
\\
\\
\underline{J}_c( 1 ) - \underline{J}_c( \pi_0 )  & \ \  \Delta_2
\end{bmatrix}^{-1}
\end{align*}
Therefore, we obtain an expression for $\mathcal{A}_2^{-1}$ as below
\begin{align}
\mathcal{A}_2^{-1}   \equiv 
\begin{bmatrix}
\displaystyle \left( (1 - \pi_0)  - \left( \int_{\pi_0}^{1} \underline{J}^{\top} (r) dr \right) \Delta_2^{-1} \big( \underline{J}_c( 1 ) - \underline{J}_c( \pi_0 ) \big) \right)^{-1} & \displaystyle - \frac{1}{1 - \pi_0} \left( \int_{\pi_0}^{1} \underline{J}^{\top} (r) dr \right) \mathcal{S}_2^{-1}
\\
\\
\displaystyle - \mathcal{S}_2^{-1} \frac{ \big( \underline{J}_c( 1 ) - \underline{J}_c( \pi_0 ) \big)}{\pi_0 (1 - \pi_0)}  \ \  & \ \  \displaystyle \mathcal{S}_2^{-1} 
\end{bmatrix}
\end{align}

\newpage 

\begin{landscape}

\begin{align*}
\mathcal{D}_1 \mathcal{Q}_{ \mathcal{R} } \mathcal{D}_2   
&=  
\bigg[ \mathcal{D}_2^{-1} \left(Z_1^{\prime} X_1 \right) \mathcal{D}_1^{-1} \bigg]^{-1}  \bigg[ \mathcal{D}_2^{-1} \left( Z_1^{\prime} Z_1 \right) \mathcal{D}_1^{-1} \bigg] \bigg[ \mathcal{D}_2^{-1} \left(Z_1^{\prime} X_1 \right) \mathcal{D}_1^{-1} \bigg]^{-1}
\nonumber
\\
&\ +  
\bigg[ \mathcal{D}_2^{-1} \left(Z_2^{\prime} X_2 \right) \mathcal{D}_1^{-1} \bigg]^{-1}  \bigg[ \mathcal{D}_2^{-1} \left( Z_2^{\prime} Z_2 \right) \mathcal{D}_1^{-1} \bigg] \bigg[ \mathcal{D}_2^{-1} \left(Z_2^{\prime} X_2 \right) \mathcal{D}_1^{-1} \bigg]^{-1}  
\\
&= 
\begin{bmatrix}
\displaystyle \left( \pi_0  - \left( \int_0^{\pi_0} \underline{J}^{\top} (r) dr \right) \Delta_1^{-1} \underline{J}_c( \pi_0 ) \right)^{-1} \ \  & \ \ \displaystyle - \frac{1}{\pi_0} \left( \int_0^{\pi_0} \underline{J}^{\top} (r) dr \right) \mathcal{S}_1^{-1}
\\
\\
\displaystyle  - \mathcal{S}_1^{-1} \frac{ \underline{J}_c( \pi_0 )}{ \pi_0 }  \ \  & \ \  \displaystyle \mathcal{S}_1^{-1} 
\end{bmatrix}
\times
\begin{bmatrix}
\pi_0 &  \ \ 0
\\
\\
\underline{J}_{c} ( \pi_0 )   &  \pi_0 V_{zz}  
\end{bmatrix}
\\
& \times
\begin{bmatrix}
\displaystyle \left( \pi_0  - \left( \int_0^{\pi_0} \underline{J}^{\top} (r) dr \right) \Delta_1^{-1} \underline{J}_c( \pi_0 ) \right)^{-1} \ \  & \ \ \displaystyle - \frac{1}{\pi_0} \left( \int_0^{\pi_0} \underline{J}^{\top} (r) dr \right) \mathcal{S}_1^{-1}
\\
\\
- \displaystyle \mathcal{S}_1^{-1} \frac{ \underline{J}_c( \pi_0 )}{ \pi_0 } \ \  & \ \  \displaystyle \mathcal{S}_1^{-1} 
\end{bmatrix}
\\
\\
&+ 
\begin{bmatrix}
\displaystyle \left( (1 - \pi_0)  - \left( \int_{\pi_0}^{1} \underline{J}^{\top} (r) dr \right) \Delta_2^{-1} \big( \underline{J}_c( 1 ) - \underline{J}_c( \pi_0 ) \big) \right)^{-1} & \displaystyle - \frac{1}{1 - \pi_0} \left( \int_{\pi_0}^{1} \underline{J}^{\top} (r) dr \right) \mathcal{S}_2^{-1}
\\
\\
\displaystyle - \mathcal{S}_2^{-1} \frac{ \big( \underline{J}_c( 1 ) - \underline{J}_c( \pi_0 ) \big)}{\pi_0 (1 - \pi_0)}  \ \  & \ \  \displaystyle \mathcal{S}_2^{-1} 
\end{bmatrix}
\begin{bmatrix}
1 - \pi_0 &  \ \ 0
\\
\\
\underline{J}_{c} ( 1 ) - \underline{J}_{c} ( \pi_0 )   &  (1 - \pi_0) V_{zz}  
\end{bmatrix}
\times
\\
&\times 
\begin{bmatrix}
\displaystyle \left( (1 - \pi_0)  - \left( \int_{\pi_0}^{1} \underline{J}^{\top} (r) dr \right) \Delta_2^{-1} \big( \underline{J}_c( 1 ) - \underline{J}_c( \pi_0 ) \big) \right)^{-1} & \displaystyle - \frac{1}{1 - \pi_0} \left( \int_{\pi_0}^{1} \underline{J}^{\top} (r) dr \right) \mathcal{S}_2^{-1}
\\
\\
\displaystyle - \mathcal{S}_2^{-1} \frac{ \big( \underline{J}_c( 1 ) - \underline{J}_c( \pi_0 ) \big)}{\pi_0 (1 - \pi_0)}  \ \  & \ \  \displaystyle \mathcal{S}_2^{-1} 
\end{bmatrix}
\end{align*}


\newpage 

\begin{align}
\Phi_1 &:= 
\begin{bmatrix}
\displaystyle \left( \pi_0  - \left( \int_0^{\pi_0} \underline{J}^{\top} (r) dr \right) \Delta_1^{-1} \underline{J}_c( \pi_0 ) \right)^{-1} \ \  & \ \ \displaystyle - \frac{1}{\pi_0} \left( \int_0^{\pi_0} \underline{J}^{\top} (r) dr \right) \mathcal{S}_1^{-1}
\\
\\
\displaystyle  - \mathcal{S}_1^{-1} \frac{ \underline{J}_c( \pi_0 )}{ \pi_0 }  \ \  & \ \  \displaystyle \mathcal{S}_1^{-1} 
\end{bmatrix}_{ \textcolor{red}{ 2 \times 2 } }
\times
\begin{bmatrix}
\pi_0 &  \ \ 0
\\
\\
\underline{J}_{c} ( \pi_0 )   &  \pi_0 V_{zz}  
\end{bmatrix}_{ \textcolor{red}{ 2 \times 2 } }
\\
&\ \times
\begin{bmatrix}
\displaystyle \left( \pi_0  - \left( \int_0^{\pi_0} \underline{J}^{\top} (r) dr \right) \Delta_1^{-1} \underline{J}_c( \pi_0 ) \right)^{-1} \ \  & \ \ \displaystyle - \frac{1}{\pi_0} \left( \int_0^{\pi_0} \underline{J}^{\top} (r) dr \right) \mathcal{S}_1^{-1}
\\
\\
- \displaystyle \mathcal{S}_1^{-1} \frac{ \underline{J}_c( \pi_0 )}{ \pi_0 } \ \  & \ \  \displaystyle \mathcal{S}_1^{-1} 
\end{bmatrix}_{ \textcolor{red}{ 2 \times 2 } }
\nonumber
\\
\nonumber
\\
&= 
\begin{bmatrix}
\displaystyle \pi_0 \left( \pi_0  - \left( \int_0^{\pi_0} \underline{J}^{\top} (r) dr \right) \Delta_1^{-1} \underline{J}_c( \pi_0 ) \right)^{-1} - \frac{1}{\pi_0} \left( \int_0^{\pi_0} \underline{J}^{\top} (r) dr \right) \mathcal{S}_1^{-1} \underline{J}_{c} ( \pi_0 ) \ \ & \ \ - \displaystyle \left( \int_0^{\pi_0} \underline{J}^{\top} (r) dr \right) \mathcal{S}_1^{-1} V_{zz}
\\
\\
- \displaystyle \mathcal{S}_1^{-1} \underline{J}_c( \pi_0 ) + \mathcal{S}_1^{-1} \underline{J}_c( \pi_0 )\  \textcolor{red}{= 0}  \ \ & \ \ \pi_0 \mathcal{S}_1^{-1} V_{zz}
\end{bmatrix}_{ \textcolor{red}{ 2 \times 2 } }
\nonumber
\\
\nonumber
\\
& \times
\begin{bmatrix}
\displaystyle \left( \pi_0  - \left( \int_0^{\pi_0} \underline{J}^{\top} (r) dr \right) \Delta_1^{-1} \underline{J}_c( \pi_0 ) \right)^{-1} \ \  & \ \ \displaystyle - \frac{1}{\pi_0} \left( \int_0^{\pi_0} \underline{J}^{\top} (r) dr \right) \mathcal{S}_1^{-1}
\\
\\
- \displaystyle \mathcal{S}_1^{-1} \frac{ \underline{J}_c( \pi_0 )}{ \pi_0 } \ \  & \ \  \displaystyle \mathcal{S}_1^{-1} 
\end{bmatrix}_{ \textcolor{red}{ 2 \times 2 } }
\nonumber
\\
\nonumber
\\
&= 
\begin{bmatrix}
\alpha_1 & \ \ \alpha_2
\\
\\
\alpha_3 & \ \ \alpha_4
\end{bmatrix}_{ \textcolor{red}{ 2 \times 2 } }
\end{align}

\newpage 

where
\begin{align*}
\alpha_1 
&= 
\left\{  \displaystyle \pi_0 \left( \pi_0  - \left( \int_0^{\pi_0} \underline{J}^{\top} (r) dr \right) \Delta_1^{-1} \underline{J}_c( \pi_0 ) \right)^{-1} - \frac{1}{\pi_0} \left( \int_0^{\pi_0} \underline{J}^{\top} (r) dr \right) \mathcal{S}_1^{-1} \underline{J}_{c} ( \pi_0 )    \right\} \left( \pi_0  - \left( \int_0^{\pi_0} \underline{J}^{\top} (r) dr \right) \Delta_1^{-1} \underline{J}_c( \pi_0 ) \right)^{-1}
\\
&+ \left( \int_0^{\pi_0} \underline{J}^{\top} (r) dr \right) \mathcal{S}_1^{-1} V_{zz} \mathcal{S}_1^{-1} \frac{ \underline{J}_c( \pi_0 )}{ \pi_0 }
\\
&= 
\displaystyle \pi_0 \left( \pi_0  - \left( \int_0^{\pi_0} \underline{J}^{\top} (r) dr \right) \Delta_1^{-1} \underline{J}_c( \pi_0 ) \right)^{-2} 
- 
\frac{1}{\pi_0} \left( \int_0^{\pi_0} \underline{J}^{\top} (r) dr \right) \mathcal{S}_1^{-1} \underline{J}_{c} ( \pi_0 ) \left( \pi_0  - \left( \int_0^{\pi_0} \underline{J}^{\top} (r) dr \right) \Delta_1^{-1} \underline{J}_c( \pi_0 ) \right)^{-1}
\\
&+ \left( \int_0^{\pi_0} \underline{J}^{\top} (r) dr \right) \mathcal{S}_1^{-1} V_{zz} \mathcal{S}_1^{-1} \frac{ \underline{J}_c( \pi_0 )}{ \pi_0 }
\end{align*}
\begin{align*}
\alpha_2 
&= 
\left\{  \displaystyle \pi_0 \left( \pi_0  - \left( \int_0^{\pi_0} \underline{J}^{\top} (r) dr \right) \Delta_1^{-1} \underline{J}_c( \pi_0 ) \right)^{-1} - \frac{1}{\pi_0} \left( \int_0^{\pi_0} \underline{J}^{\top} (r) dr \right) \mathcal{S}_1^{-1} \underline{J}_{c} ( \pi_0 )    \right\} \left\{  - \frac{1}{\pi_0} \left( \int_0^{\pi_0} \underline{J}^{\top} (r) dr \right) \mathcal{S}^{-1} \right\}
\\
&- \left( \int_0^{\pi_0} \underline{J}^{\top} (r) dr \right) \mathcal{S}_1^{-1} V_{zz} \mathcal{S}_1^{-1} 
\\
&= 
\displaystyle - \left( \pi_0  - \left( \int_0^{\pi_0} \underline{J}^{\top} (r) dr \right) \Delta_1^{-1} \underline{J}_c( \pi_0 ) \right)^{-1} \left( \int_0^{\pi_0} \underline{J}^{\top} (r) dr \right) \mathcal{S}_1^{-1} 
+ \frac{1}{\pi_0^2} \left( \int_0^{\pi_0} \underline{J}^{\top} (r) dr \right) \mathcal{S}_1^{-1} \underline{J}_{c} ( \pi_0 ) \left( \int_0^{\pi_0} \underline{J}^{\top} (r) dr \right) \mathcal{S}^{-1}  
\\
&- \left( \int_0^{\pi_0} \underline{J}^{\top} (r) dr \right) \mathcal{S}_1^{-1} V_{zz} \mathcal{S}_1^{-1}
\end{align*}
\begin{align*}
\alpha_3 &= - S_1^{-1} V_{zz} S_1^{-1} \underline{J}_c ( \pi_0 )   
\\
\\
\alpha_4 &= \pi_0 S_1^{-1} V_{zz} S_1^{-1}    
\end{align*}

\newpage 

\begin{align*}
\Phi_2 
&:= 
\begin{bmatrix}
\displaystyle \left( (1 - \pi_0)  - \left( \int_{\pi_0}^{1} \underline{J}^{\top} (r) dr \right) \Delta_2^{-1} \big( \underline{J}_c( 1 ) - \underline{J}_c( \pi_0 ) \big) \right)^{-1} & \displaystyle - \frac{1}{1 - \pi_0} \left( \int_{\pi_0}^{1} \underline{J}^{\top} (r) dr \right) \mathcal{S}_2^{-1}
\\
\\
\displaystyle - \mathcal{S}_2^{-1} \frac{ \big( \underline{J}_c( 1 ) - \underline{J}_c( \pi_0 ) \big)}{\pi_0 (1 - \pi_0)}  \ \  & \ \  \displaystyle \mathcal{S}_2^{-1} 
\end{bmatrix}
\begin{bmatrix}
1 - \pi_0 &  \ \ 0
\\
\\
\underline{J}_{c} ( 1 ) - \underline{J}_{c} ( \pi_0 )   &  (1 - \pi_0) V_{zz}  
\end{bmatrix}
\\
&\ \times 
\begin{bmatrix}
\displaystyle \left( (1 - \pi_0)  - \left( \int_{\pi_0}^{1} \underline{J}^{\top} (r) dr \right) \Delta_2^{-1} \big( \underline{J}_c( 1 ) - \underline{J}_c( \pi_0 ) \big) \right)^{-1} & \displaystyle - \frac{1}{1 - \pi_0} \left( \int_{\pi_0}^{1} \underline{J}^{\top} (r) dr \right) \mathcal{S}_2^{-1}
\\
\\
\displaystyle - \mathcal{S}_2^{-1} \frac{ \big( \underline{J}_c( 1 ) - \underline{J}_c( \pi_0 ) \big)}{\pi_0 (1 - \pi_0)}  \ \  & \ \  \displaystyle \mathcal{S}_2^{-1} 
\end{bmatrix}
\\
&= 
\begin{bmatrix}
\displaystyle \left\{ (1 - \pi_0) \left( (1 - \pi_0)  - \left( \int_{\pi_0}^{1} \underline{J}^{\top} (r) dr \right) \Delta_2^{-1} \big( \underline{J}_c( 1 ) - \underline{J}_c( \pi_0 ) \big)  \right)^{-1} - \frac{1}{1- \pi_0} \left( \int_{\pi_0}^{1} \underline{J}^{\top} (r) dr \right) \mathcal{S}_2^{-1} \big( \underline{J}_c( 1 ) - \underline{J}_c( \pi_0 ) \big) \right\} & - \displaystyle \left( \int_0^{\pi_0} \underline{J}^{\top} (r) dr \right) \mathcal{S}_2^{-1} V_{zz}
\\
\\
- \displaystyle \mathcal{S}_2^{-1} \big( \underline{J}_c( 1 ) - \underline{J}_c( \pi_0 ) \big) + \mathcal{S}_2^{-1} \big( \underline{J}_c( 1 ) - \underline{J}_c( \pi_0 ) \big) \  \textcolor{red}{= 0}  &  (1 - \pi_0 ) \mathcal{S}_2^{-1} V_{zz}
\end{bmatrix}
\nonumber
\\
\nonumber
\\
& \times
\begin{bmatrix}
\displaystyle \left( (1 - \pi_0)  - \left( \int_{\pi_0}^{1} \underline{J}^{\top} (r) dr \right) \Delta_2^{-1} \big( \underline{J}_c( 1 ) - \underline{J}_c( \pi_0 ) \big) \right)^{-1} & \displaystyle - \frac{1}{1 - \pi_0} \left( \int_{\pi_0}^{1} \underline{J}^{\top} (r) dr \right) \mathcal{S}_2^{-1}
\\
\\
\displaystyle - \mathcal{S}_2^{-1} \frac{ \big( \underline{J}_c( 1 ) - \underline{J}_c( \pi_0 ) \big)}{\pi_0 (1 - \pi_0)}  \ \  & \ \  \displaystyle \mathcal{S}_2^{-1} 
\end{bmatrix}
\nonumber
\\
\nonumber
\\
&= 
\begin{bmatrix}
\beta_1 & \ \ \beta_2
\\
\\
\beta_3 & \ \ \beta_4
\end{bmatrix}_{ \textcolor{red}{ 2 \times 2 } }
\end{align*}

\end{landscape}

\newpage 

\section{Supplementary Material}

Consider the predictive regression model given by
\begin{align}
y_t &= \beta x_{t-1} + u_t \\
x_t &= \left( 1 - \frac{c}{n} \right) x_{t-1} + v_t
\end{align}
Consider the asymptotic distribution of the term $\frac{1}{n^{1 + \delta}} \sum_{t=1}^n x_{t-1} \tilde{z}_{t-1}$. Following KMS we have the following expression
\begin{align}
\frac{1}{n^{1 + \delta}} \sum_{t=1}^n x_{t-1} \tilde{z}_{t-1} = \frac{1}{n^{1 + \delta}} \sum_{t=1}^n x_{t-1} z_{t-1} - \frac{c}{c_z}  \sum_{t=1}^n \frac{x^2_{t-1}}{ n^2 } + o_p(1)
\end{align}
Also, from PM09 we obtain that
\begin{align}
\frac{1}{n^{1 + \delta}} \sum_{t=1}^n x_{t-1} z_{t-1} = - \frac{1}{c_z} \left[ \sum_{t=1}^n \frac{x_{t-1} v_{t}}{n} + \sum_{t=1}^n \frac{z_{t-1} v_{t}}{n} + \sum_{t=1}^n \frac{v^2_{t}}{n} \right] + o_p(1)
\end{align}
Therefore, we obtain that
\begin{align}
\frac{1}{n^{1 + \delta}} \sum_{t=1}^n x_{t-1} \tilde{z}_{t-1} = - \frac{1}{c_z} \left[ \sum_{t=1}^n \frac{x_{t-1} v_{t}}{n} + \sum_{t=1}^n \frac{z_{t-1} v_{t}}{n} + \sum_{t=1}^n \frac{v^2_{t}}{n} + c \sum_{t=1}^n \frac{x^2_{t-1}}{ n^2 } \right] + o_p(1)
\end{align}
and the asymptotic convergence result follows, since by PM when $x_t$ is a local to unity stochastic process, then $\sum_{t=1}^n \frac{z_{t-1} v_{t}}{n} \to \Lambda_{xx}$.
\begin{align}
\frac{1}{n^{1 + \delta}} \sum_{t=1}^n x_{t-1} \tilde{z}_{t-1} &\Rightarrow - \frac{1}{c_z} \left[ \int_0^1 J_c dB_v + \sigma_v^2 + c\int_0^1 J_c^2  \right]
\nonumber
\\
&\equiv - \frac{1}{c_z} \left[ \sigma_v^2 + \int_0^1 J_c \bigg( dB_v + c J_c \bigg)  \right]
\nonumber
\\
&=  - \frac{1}{c_z} \left[ \sigma_v^2 + \int_0^1 J_c dJ_c \right]
\nonumber
\\
&:= - \frac{1}{c_z}  Q(1).
\end{align}

The above result also implies that in terms the post-break and pre-break sample moments we obtain the following weakly convergence of the corresponding limit
\begin{align}
\frac{1}{n^{1 + \delta}} \sum_{t=1}^k x_{t-1} \tilde{z}_{t-1} &\Rightarrow - \frac{1}{c_z} \left[ \pi \sigma_v^2 + \int_0^{\pi} J_c dJ_c \right]
:= - \frac{1}{c_z}  Q( \pi ),
\\
\frac{1}{n^{1 + \delta}} \sum_{t=k+1}^n x_{t-1} \tilde{z}_{t-1} &\Rightarrow - \frac{1}{c_z} \left[ (1 - \pi) \sigma_v^2 + \int_{\pi}^{1} J_c dJ_c \right]
:= - \frac{1}{c_z} \big( Q( 1 ) - Q( \pi )  \big) .
\end{align}

\newpage 

Therefore, considering the Wald-IVX statistic for the univariate predictive regression with a single regressor and no model intercept,  we obtain the following expression
\begin{align}
\mathcal{W}_T^{\text{IVX}} ( \pi ) = \frac{ \left\{ \left( \sum_{t=1}^T  Z_1 u \right) \left( \sum_{t=1}^T Z X \right) - \left( \sum_{t=1}^T  Z u \right) \left( \sum_{t=1}^T Z_1 X_1 \right) \right\}^2 }{  \hat{\sigma}_u^2 \mathcal{I} }
\end{align}
where 
\begin{align}
\mathcal{I} = \left\{ \left( \sum_{t=1}^T Z_1^2 \right) \left( \sum_{t=1}^T Z X \right)^2 - 2 \left( \sum_{t=1}^T Z_1^2 \right) \left( \sum_{t=1}^T Z X \right) \left( \sum_{t=1}^T Z_1 X_1 \right) +  \left( \sum_{t=1}^T Z_1 X_1 \right)^2 \left( \sum_{t=1}^T Z^2 \right)  \right\} 
\end{align}
The above simplifies to the following expression 
\begin{align}
\label{interesting}
\mathcal{W}_T^{\text{IVX}} ( \pi ) =  \frac{ \left\{ \displaystyle \left( \sum_{t=1}^T  Z_{1t} u_{t+1} \right) - \displaystyle \frac{ \sum_{t=1}^T Z_{1t} X_{1t} }{ \sum_{t=1}^T Z_{t} X_{t} } \left( \sum_{t=1}^T  Z_{t} u_{t+1} \right) \right\}^2 }{ \displaystyle \hat{\sigma}_u^2 \left\{ \sum_{t=1}^T Z_{1t}^2 - 2 \left( \sum_{t=1}^T Z_{1t}^2 \right) \left( \frac{ \sum_{t=1}^T Z_{1t} X_{1t} }{  \sum_{t=1}^T Z_t X_t } \right) + \sum_{t=1}^T Z_t^2 \left( \frac{ \sum_{t=1}^T Z_{1t} X_{1t} }{ \sum_{t=1}^T Z_t X_t } \right)^2  \right\} }
\end{align}
We use $Z_2 = \left( Z - Z_1 \right)$ and $X_2 = \left( X - X_1 \right)$, which also implies that $\left( Z_2 X_2 \right)= \left( ZX - Z X_1 \right)$ and also we have that $(Z X_1) = (Z_1 X_1)$ and $(Z Z_1) = (Z_1^2)$.

Furthermore, by defining 
\begin{align}
w_t = Z_{1t} - \left( \frac{ \sum_{t=1}^T Z_{1t} X_{1t} }{ \sum_{t=1}^T Z_t X_t } \right) Z_t    
\end{align}
Thus, we obtain that
\begin{align}
\label{main.expression}
\mathcal{W}_T^{\text{IVX}} ( \pi )  =  \frac{1}{ \hat{\sigma}^2_u} \frac{ \displaystyle \left\{ \sum_{t=1}^T w_t u_{t+1} \right\}^2 }{ \displaystyle  \left( \sum_{t=1}^T w^2_t \right) } 
\end{align}

Note that expression \eqref{main.expression}, provides an alternative representation of the Wald IVX statistic which make it easier to check the main theoretical results of the paper.

\newpage 

\underline{Case I:} known break-point (single persistent predictor)

In particular, in the case of a known break-point, say $\pi \equiv \pi_0$ we can see that 
\begin{align}
\mathcal{W}_T^{\text{IVX}} ( \pi_0 )  =  \frac{1}{\hat{\sigma}^2_u} \frac{ \displaystyle \left\{ \sum_{t=1}^T w_t u_{t+1} \right\}^2 }{ \displaystyle  \left( \sum_{t=1}^T w^2_t \right) } 
\to \frac{1}{\sigma^2_u} \frac{ \bigg\{ \mathcal{N} \bigg( 0, \sigma^2_u   W(\pi_0 ) \bigg)   \bigg\}^2  }{   W(\pi_0 ) } \equiv \chi^2(1).
\end{align}
where $\hat{\sigma}^2_u$ a consistent estimator of $\sigma^2$.

\begin{proof}
Note that using similar arguments (only the decomposition) as in Section \eqref{additional}, one can show that 
\begin{align}
\sum_{t=1}^T \frac{ w_t u_{t+1} }{ n^{1 + \delta} }  \overset{ d }{ \to } 
\mathcal{MN} \bigg( 0, \Phi_c ( \pi_0 ) \sigma_u^2 \bigg) 
\end{align}
and that 
\begin{align}
\sum_{t=1}^T \frac{w^2_t}{ n^{1 + \delta} }  \overset{ p }{ \to } 
\Phi_c ( \pi_0 ) 
\end{align}
Therefore, by a simply application of the continuous mapping theorem by substituting in the expression for the Wald IVX statistic (in the case of a known break-point) then we obtain
\begin{align}
\mathcal{W}_T^{\text{IVX}} ( \pi_0 )  \overset{ d }{ \to }  \frac{1}{ \sigma^2_u} \frac{ \displaystyle \left\{\mathcal{MN} \bigg( 0, \Phi_c ( \pi_0 ) \sigma_u^2 \bigg) \right\}^2 }{ \displaystyle   \Phi_c ( \pi_0 ) } 
\equiv \chi^2(1).
\end{align}
which is a $\chi^2(1)$ random variable with 1 degree of freedom.

\end{proof}

\newpage

\bibliographystyle{apalike}
\bibliography{myreferences1}

\end{document}